\documentclass[particles,article,accept,moreauthors,pdftex]{mdpi} 
\firstpage{1} 
\pubvolume{xx}
\issuenum{1}
\articlenumber{5}
\pubyear{2020}
\copyrightyear{2020}
\history{Received: 15 January 2020; Accepted: 24 February 2020; Published: date}
\updates{yes} 

\Title{Exploring the Partonic Phase at Finite
Chemical Potential in and out-of Equilibrium}

\Author{O. Soloveva$^{1}$, P. Moreau $^{1,2}$, L. Oliva $^{1}$, 
V. Voronyuk $^{3,4}$, V. Kireyeu $^{3}$, T. Song $^{5}$ and E.~Bratkovskaya $^{1,5,*}$}
\AuthorNames{Firstname Lastname, Firstname Lastname and Firstname Lastname}

\address{%
$^{1}$ \quad Institut f\"ur Theoretische Physik,  Goethe-Universit\"{a}t Frankfurt am Main, Frankfurt am Main, Germany; soloveva@fias.uni-frankfurt.de (O.S.); pierre.moreau@duke.edu (P.M.); oliva@fias.uni-frankfurt.de (L.O.) \\
$^{2}$ \quad Department of Physics, Duke University, Durham, NC 27708, USA \\
$^{3}$ \quad  Joint Institute for Nuclear Research, Joliot-Curie 6, 141980 Dubna, Moscow Region, Russia; vadim.voronyuk@jinr.ru (V.V.); vkireyeu@jinr.ru (V.K.) \\
$^{4}$ \quad Institute for Theoretical Physics, Metrolohichna str. 14-b, 03143 Kiev, Ukraine \\
$^{5}$ \quad GSI Helmholtzzentrum f\"{u}r Schwerionenforschung GmbH, Darmstadt, Germany; T.Song@gsi.de}

\corres{Correspondence: E.Bratkovskaya@gsi.de }

\abstract{
We study the influence of the baryon chemical potential $\mu_B$  on the properties
of the Quark--Gluon--Plasma (QGP) in and out-of equilibrium. The description of the
QGP in equilibrium is based on the effective propagators and couplings from the Dynamical QuasiParticle Model (DQPM) that is matched to reproduce the equation-of-state of the partonic system above the deconfinement temperature $T_c$ from lattice QCD.
We study the transport coefficients such as the ratio of shear viscosity $\eta$
and  bulk viscosity $\zeta$ over entropy density $s$, i.e., $\eta/s$ and $\zeta/s$ 
in the $(T,\mu)$ plane and compare to other model results available at $\mu_B =0$.
The out-of equilibrium study of the QGP is performed within 
the Parton--Hadron--String Dynamics (PHSD) transport approach extended in the 
partonic sector by explicitly calculating the total and differential partonic 
scattering cross sections based on the DQPM and the evaluated 
at actual temperature $T$ and baryon chemical potential $\mu_B$ in each 
individual space-time cell where partonic scattering takes place.
The traces of their $\mu_B$ dependences are investigated in 
different observables for symmetric Au+Au and asymmetric Cu+Au collisions 
such as rapidity and $m_T$-distributions and directed and elliptic flow 
coefficients $v_1, v_2$ in the energy range 7.7 GeV $\le \sqrt{s_{NN}}\le 200$ GeV. }

\begin{document}

\section{Introduction}

The phase diagram of matter is one of the most fascinating subjects in physics, which 
also has important implications on chemistry and biology. Its phase boundaries and (possibly) critical points have been the focus of physics research for centuries. 
Apart from the traditional phase diagram in the plane of temperature $T$ and pressure $P$,
its transport properties like the shear and bulk viscosities, the~electric conductivity, etc. are also of fundamental interest. These transport coefficients emerge from the stationary limit of correlators and provide additional information on the systems in thermal and chemical equilibrium apart from the equation of state.
In this context, the~phase diagram of strongly interacting matter has been the topic of most interest for decades and substantial experimental and theoretical efforts have been invested to shed light on this issue. It contains the information about the properties of our universe from early beginning---directly after the Big Bang---when the matter was in the QGP phase at very large temperature $T$ and about zero baryon chemical potential $\mu_B$, 
to the later stages of the universe, where in the expansion phase stars and Galaxy have been formed. Here, the matter is at low temperature and large baryon chemical potential.
Relativistic and ultra-relativistic heavy-ion collisions (HICs) nowadays offer the unique possibility to study some of these phases, in~particular a QGP phase and its phase 
boundary to the hadronic one. Furthermore, the~phase diagram of strongly interacting matter in the $(T, \mu_B$) plane can also be explored in the astrophysical context at moderate temperatures and high $\mu_B$ \cite{Klahn:2006ir}, i.e.,~in the dynamics of supernovae or---more recently---in the dynamics of neutron star~merges. 

In order to reproduce the mini Big Bangs in laboratories,  heavy-ion accelerators 
are built  which allow for investigating the creation of the QGP under controlled conditions.
Hadronic spectra and relative hadron abundances from these experiments
reflect  important aspects of the dynamics in the hot and dense zone
formed in the early phase of the reaction and collective flows provide 
information on the transport properties of the medium generated on short time scales.
Whereas heavy-ion reactions at Relativistic Heavy-Ion Collider (RHIC) 
and Large Hadron Collider (LHC) energies probe a partonic medium at small baryon chemical potential $\mu_B$, the current interest is in collisions at lower bombarding energies where the net baryon density is higher and $\mu_B$ accordingly. Such conditions will be realized in future accelerators at the Facility for Antiproton and Ion Research (FAIR) in Darmstadt and the Nuclotron-based Ion Collider fAcility (NICA) in~Dubna.

Current methods to explore QCD in Minkowski space for non-vanishing
quark (or baryon) densities (or chemical potential) are effective approaches. 
Using effective models, one can 
study the dominant properties of QCD in equilibrium, i.e.,~thermodynamic quantities
as well as transport coefficients. To~this aim, the dynamical quasiparticle model (DQPM)
has been introduced~\cite{Peshier:2005pp,Cassing:2007nb,Cassing:2007yg,Linnyk:2015rco,Hamsa:JModPhys16},
which is based on partonic propagators with sizeable
imaginary parts of the self-energies incorporated. Whereas the real part of
the self-energies can be attributed to a dynamically generated mass (squared), the
imaginary parts contain the information about the interaction rates in the system.
Furthermore, the~imaginary parts of the propagators define the spectral functions of
the degrees of freedom which might show narrow (or broad) quasiparticle peaks. 
A further advantage of a propagator based approach is that one can formulate a
consistent thermodynamics~\cite{Baym} as well as a causal theory for non-equilibrium
configurations on the basis of Kadanoff--Baym equations~\cite{KadanoffBaym}. 

In order to explore the properties of the QGP close to equilibrium,  transport coefficients are calculated such as shear  $\eta$ and bulk $\zeta$ viscosities, electric conductivity 
$\sigma_0$, etc. 
While basically all of the effective models have similar equations of state (EoS), which match well with available lattice QCD data at $\mu_B =0$, the~transport coefficients can vary significantly for different models (cf.~\cite{Marty:NJL13}).
Exploration of transport coefficients of the hot and dense QGP can provide useful information for simulations of heavy-ion collisions (HIC) based on hydrodynamical models for which they 
are used as input parameters.
The experimental data for elliptic flow can be well reproduced by hydrodynamical simulations with a small value for the shear viscosity over entropy density~\cite{Romatschke,Song:Heinz}. 

Since relativistic heavy-ion collisions start with impinging nuclei in their groundstates, a proper non-equilibrium description of the entire dynamics through possibly different phases up to the final asymptotic hadronic states---eventually showing some degree of equilibration---is mandatory. To~this aim, the Parton--Hadron--String Dynamics (PHSD) transport approach
~\cite{Cassing:2008sv,Cassing:2008nn,Cassing:2009vt,Bratkovskaya:2011wp,Linnyk:2015rco}
has been formulated more then a decade ago (on the basis of the 
Hadron-String-Dynamics (HSD) approach~\citep{Cassing:1999es}),
and it was found to well describe observables from p+A and A+A collisions from SPS to LHC energies including electromagnetic probes such as photons and dileptons 
\citep{Linnyk:2015rco}.
In order to explore the partonic systems at higher $\mu_B$, the PHSD approach has been recently extended to incorporate partonic quasiparticles and their differential cross sections that depend not only on temperature $T$ as in the previous PHSD studies, but~
also on  chemical potential $\mu_B$ explicitly---cf.~\cite{Moreau:2019vhw}.
Within this extended approach, we have studied the `bulk' observables  in HIC for different
energies---from AGS to RHIC,  and~systems---strongly asymmetric C+Au and 
symmetric Au+Au/Pb+Pb collisions. We have found only a small influence of $\mu_B$ dependences of parton properties (masses and widths) and their interaction 
cross sections in bulk observables~\cite{Moreau:2019vhw}.

In this work, we extend our study from Ref.~\cite{Moreau:2019vhw} to the 
collective flow ($v_1$, $v_2$) coefficients and their sensitivity to 
the $\mu_B$ dependences of partonic cross sections. In addition, we explore the relations 
between the in and out-of equilibrium QGP by means of transport coefficients 
and collective flows.
Additionally, we show explicitly the `bulk' results for asymmetric heavy-ion collisions 
such as Cu+Au and discuss which hadronic species and observables are more sensitive 
to such~effects.

\section{The PHSD~Approach}

We start with reminding the basic ideas of the PHSD transport approach and the DQPM.
The~Parton--Hadron--String Dynamics  transport approach \cite{Cassing:2008sv,Cassing:2008nn,Cassing:2009vt,Bratkovskaya:2011wp,Linnyk:2015rco}
is a microscopic off-shell transport approach for the description of strongly interacting hadronic and partonic matter in and out-of equilibrium. It is based on the solution of
Kadanoff--Baym equations in first-order gradient \mbox{expansion
\citep{Cassing:2008nn}} employing `resummed' propagators from the dynamical
quasiparticle model (DQPM)~\cite{Peshier:2005pp,Cassing:2007nb,Cassing:2007yg}
 for the partonic~phase.

The Dynamical Quasiparticle Model (DQPM) has been introduced in Refs.~\cite{Peshier:2005pp,Cassing:2007nb,Cassing:2007yg} for the effective description
of the properties of the QGP in terms of strongly interacting quarks and gluons
with properties and interactions which 
are adjusted to reproduce lQCD results
on the thermodynamics of the equilibrated QGP at finite temperature $T$ and 
baryon (or quark) chemical potential $\mu_q$.
In~the DQPM, the quasiparticles are characterized by single-particle Green's functions
(in propagator representation) with complex self-energies. 
The real part of the self-energies is related to the mean-field properties, 
whereas the imaginary part provides information about the lifetime and/or 
reaction rates of the~particles.

In PHSD, the partons (quarks and gluons) are strongly interacting quasiparticles 
characterized by broad spectral functions $\rho_j$ ($j=q, {\bar q}, g$), i.e.,~ 
they are off-shell contrary to the conventional cascade or transport models 
dealing with on-shell particles, i.e.,~the $\delta$-functions in the invariant mass squared. 
The quasiparticle spectral functions have a Lorentzian form~\cite{Linnyk:2015rco}
and depend on the parton mass and width parameters:
\begin{eqnarray}
\!\!\!\!\!\! \rho_j(\omega,{\bf p}) =
 \frac{\gamma_j}{E_j} \left(
   \frac{1}{(\omega-E_j)^2+\gamma_j^2} - \frac{1}{(\omega+E_j)^2+\gamma_j^2}
 \right)\
\label{eq:rho}
\end{eqnarray}
separately for quarks/antiquarks and gluons ($j=q,\bar{q},g$). With~the convention $E^2({\bf p}^2) = {\bf p}^2+M_j^2-\gamma_j^2$, the~parameters $M_j^2$ and $\gamma_j$ are directly related to the real
and imaginary parts of the retarded self-energy, e.g.,~$\Pi_j =
M_j^2-2i\gamma_j\omega$.

\begin{figure}[H]
\centerline{ \includegraphics[width=75mm]{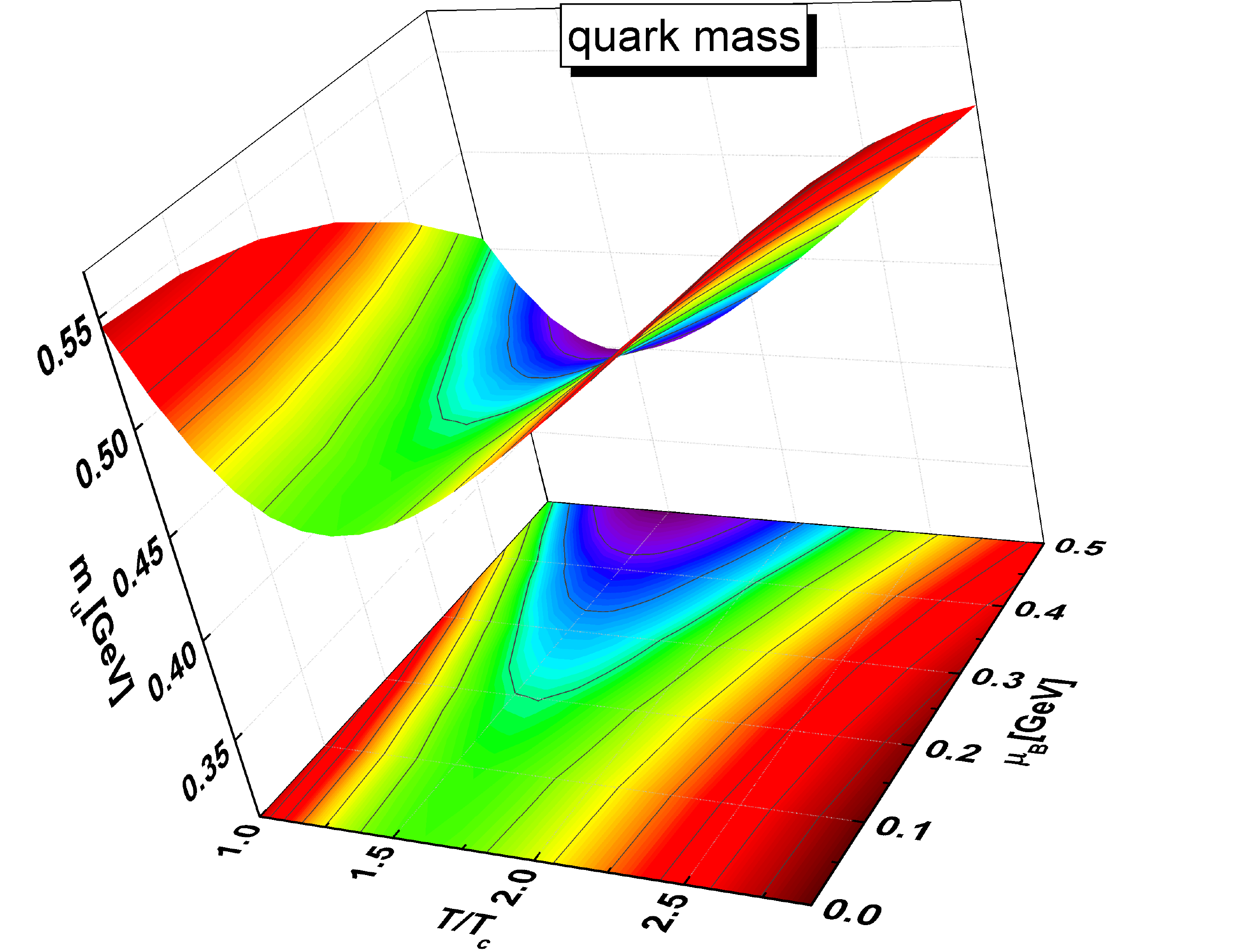}
\includegraphics[width=75mm]{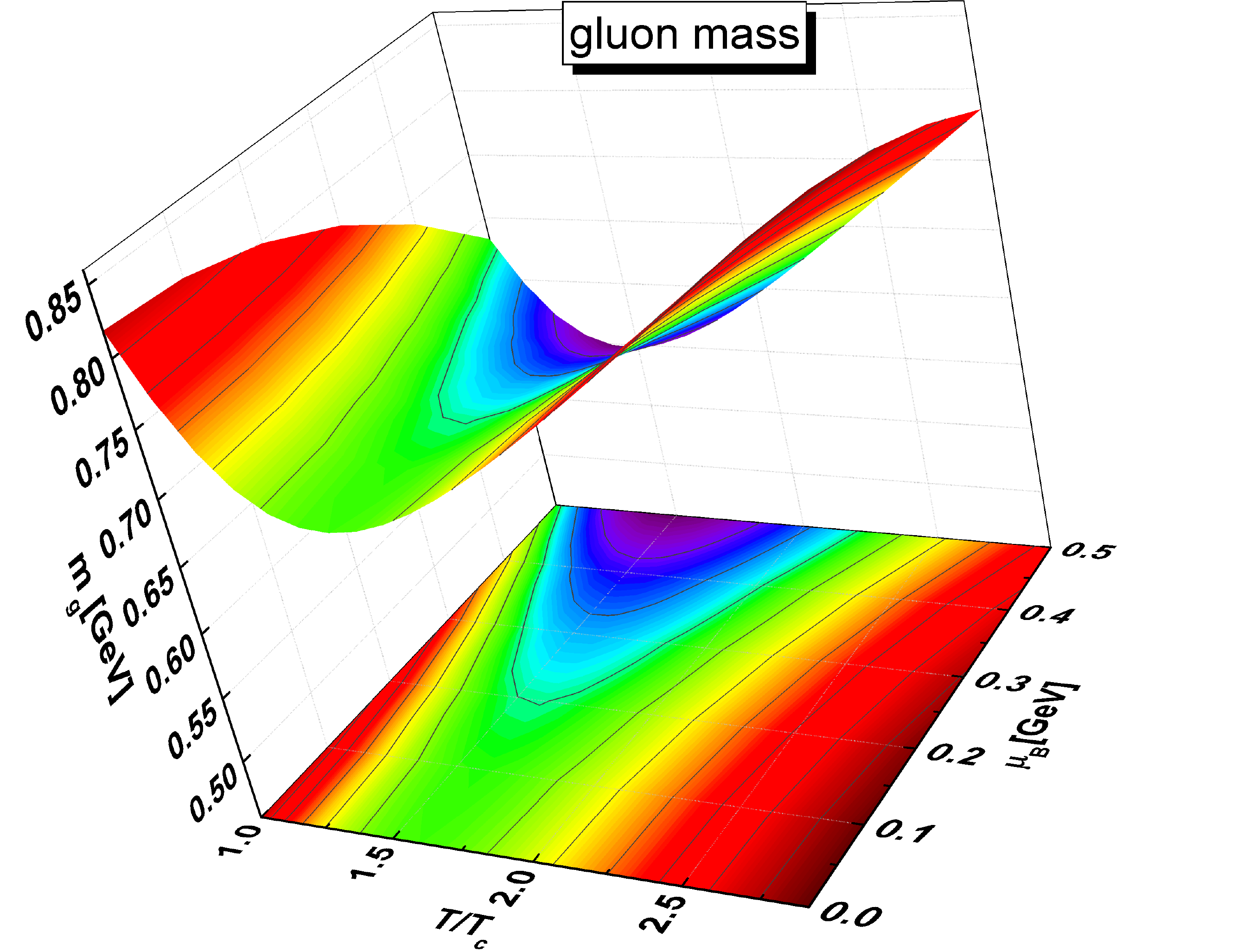}}
\centerline{ \includegraphics[width=75mm]{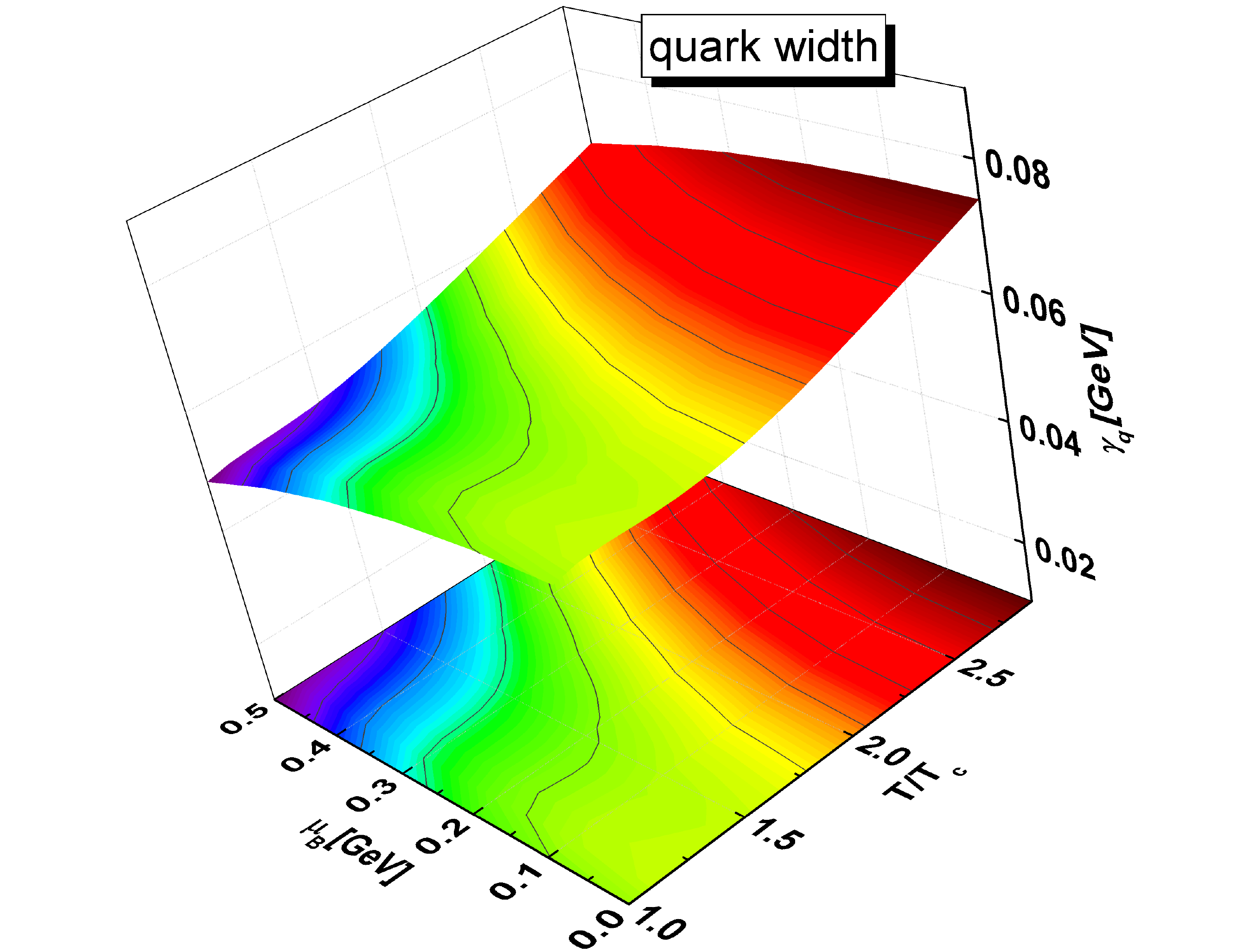}
\includegraphics[width=75mm]{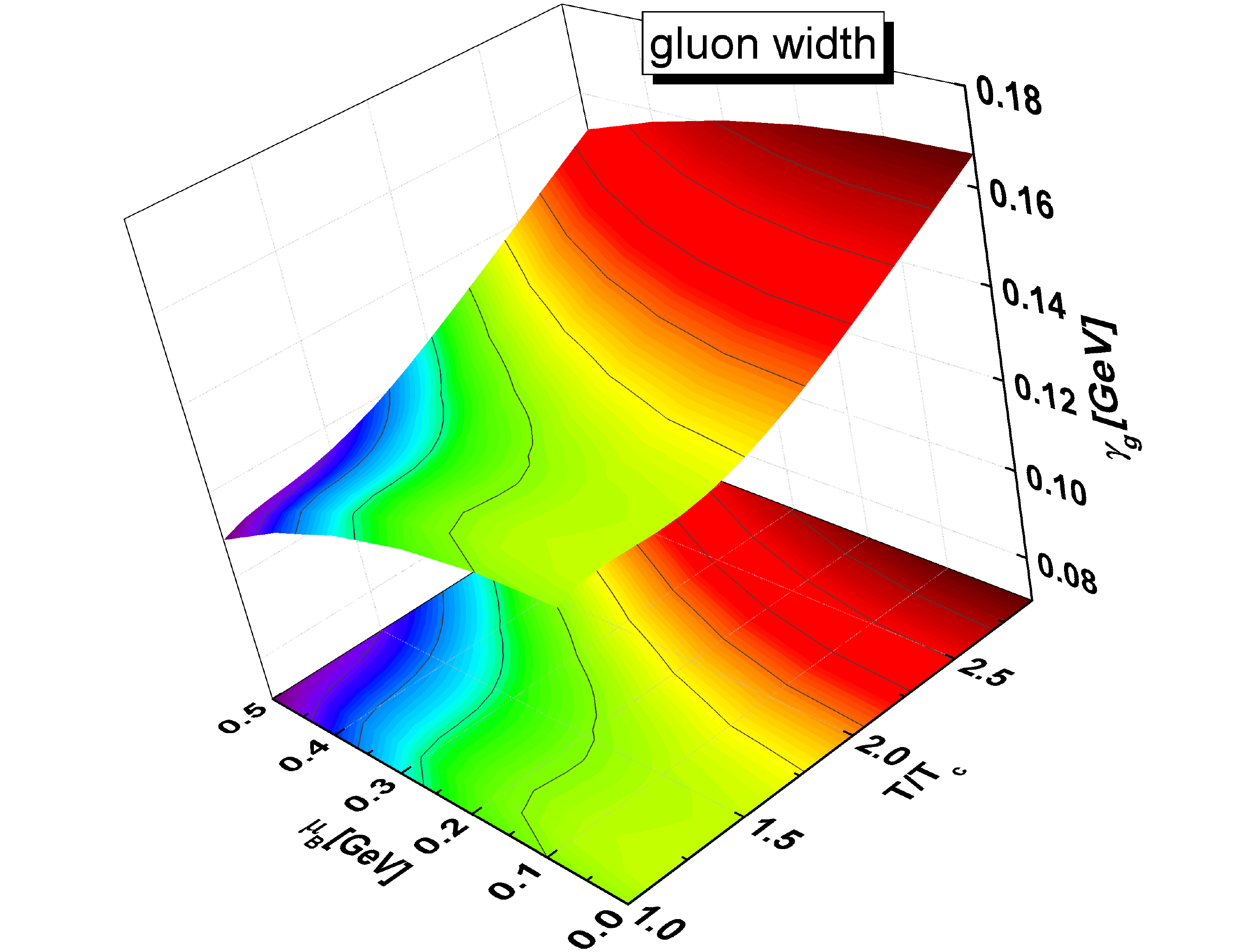}}
\caption{The effective quark (\textbf{left}) and gluon (\textbf{right}) masses $M$ 
(\textbf{upper row}) and widths $\gamma$ (\textbf{lower row}) 
 as a function of the temperature $T$ and baryon chemical potential $\mu_B$. }
\label{fig1}
\end{figure}

The actual parameters in Equation~(\ref{eq:rho}),  i.e.,~the gluon
mass $M_g$ and width $\gamma_g$---employed as input in the present PHSD
calculations---as well as the quark mass $M_q$ and width
$\gamma_q$, are depicted in Figure~\ref{fig1} as a function of the
 temperature $T$ and  baryon chemical potential $\mu_B$. These values for the masses and widths
have been fixed by fitting the lattice QCD results from Ref.~\cite{Borsanyi:2012cr,Borsanyi:2013bia} in thermodynamic equilibrium.
One can see that the masses of quarks and gluons decrease with 
increasing $\mu_B$, and a similar trend holds for the widths of~partons.

A scalar mean-field $U_s(\rho_s)$ for quarks and antiquarks can be
defined by the derivative of the potential energy density with respect to the 
scalar density $\rho_s$,
\begin{equation} \label{uss}
U_s(\rho_s) = \frac{d V_p(\rho_s)}{d \rho_s} ,
\end{equation}
which is evaluated numerically within the DQPM. Here, the potential energy density
is defined by
\begin{equation} \label{Vp}
V_p(T,\mu_q) = T^{00}_{g-}(T,\mu_q) + T^{00}_{q-}(T,\mu_q) + T^{00}_{{\bar q}-}(T,\mu_q),
\end{equation}
where the different contributions $T^{00}_{j-}$ correspond to the
space-like part of the energy-momentum tensor component $T^{00}_{j}$
of parton $j = g, q, \bar{q}$ (cf. Section~3 in Ref.
\citep{Cassing:2007nb}). The~scalar mean-field $U_s(\rho_s)$ for quarks
and antiquarks is repulsive as a function of the parton scalar
density $\rho_s$ and shows that the scalar mean-field potential is in the
order of a few GeV for $\rho_s > 10$ fm$^{-3}$. The~mean-field potential
(\ref{uss}) is employed in the PHSD transport calculations and
determines the force on a partonic quasiparticle $j$, i.e.,~$ \sim
M_j/E_j \nabla U_s(x) = M_j/E_j \ d U_s/d \rho_ s \ \nabla
\rho_s(x)$, where the scalar density $\rho_s(x)$ is determined
numerically on a space-time~grid. 

Furthermore, a~two-body interaction strength can be extracted from 
the DQPM as well from the quasiparticle width in line with Ref.~\citep{Peshier:2005pp}. 
On the partonic side, the following elastic and inelastic
interactions are included in the latest version of PHSD (v. 5.0)  
$qq \leftrightarrow qq$, $\bar{q} \bar{q} \leftrightarrow \bar{q}\bar{q}$, $gg \leftrightarrow gg$,
$gg \leftrightarrow g$, $q\bar{q} \leftrightarrow g$, $q g \leftrightarrow q g$,
$g \bar{q} \leftrightarrow g \bar{q}$  exploiting
'detailed-balance' with cross sections calculated from the leading order Feynman diagrams employing the effective propagators and couplings $g^2(T/T_c)$ from the DQPM~\cite{Moreau:2019vhw}. In~Ref.~\cite{Moreau:2019vhw}, the differential and total off-shell 
cross sections have been evaluated as a function of the invariant energy of 
colliding off-shell partons $\sqrt{s}$ for each $T$, $\mu_B$. 
We remind that in the previous PHSD studies (using v. 4.0 and below) the cross sections
depend only on $T$ as evaluated in Ref.~\cite{Ozvenchuk13}.

When implementing the differential cross sections and parton masses into the PHSD5.0
approach, one has to specify the 'Lagrange parameters' $T$ and $\mu_B$ in each 
computational cell in space-time. This has been done by employing the lattice equation 
of state and a diagonalization of the energy-momentum tensor from PHSD as described in 
Ref. \citep{Moreau:2019vhw}.

The transition from partonic to hadronic d.o.f. (and vice~versa)  is described by
covariant transition rates for the fusion of quark--antiquark pairs
or three quarks (antiquarks), respectively, obeying flavor
current--conservation, color neutrality as well as energy--momentum
conservation~\citep{Cassing:2009vt}. Since the dynamical quarks and
antiquarks become very massive close to the phase transition, the~formed resonant 'prehadronic' color-dipole states ($q\bar{q}$ or
$qqq$) are of high invariant mass, too, and~sequentially decay to
the groundstate meson and baryon octets, thus increasing the total
entropy.

On the hadronic side, PHSD includes explicitly the baryon octet and
decouplet, the~$0^-$- and $1^-$-meson nonets as well as selected
higher resonances as in the Hadron--String--Dynamics (HSD)
approach~\citep{Cassing:1999es}.  Note that PHSD and HSD
(without explicit partonic degrees-of-freedom) merge at low energy
density, in~particular below the local critical energy density
$\varepsilon_c\approx$ 0.5~GeV/fm$^{3}$.

\section{Transport~Coefficients}

The transport properties of the QGP close to equilibrium can be characterized by 
various transport coefficients. 
The shear viscosity $\eta$ and bulk viscosity $\zeta$ describe the fluid's 
dissipative corrections at leading order. Both coefficients are generally expected
to depend on the temperature $T$ and baryon chemical potential $\mu_B$. 
In the hydrodynamic equations, the~viscosities appear as dimensionless ratios, 
$\eta/s$ and $\zeta/s$, where $s$ is the fluid entropy density. 
Such specific viscosities are more meaningful than the unscaled $\eta$ and $\zeta$
values because~they describe the magnitude of stresses inside
the medium relative to its natural~scale. 

In our recent studies~\cite{Moreau:2019vhw,Soloveva:2019xph,Soloveva:2019doq}, we have 
investigated the transport properties of the QGP in the $(T,\mu_B)$ plane based on the DQPM.
One way to evaluate the viscosity coefficients of partonic matter is the Kubo formalism~\cite{Kubo,Aarts:2002,Fraile:2006,Lang12}, which was used to calculate the viscosities for a previous version of the DQPM within the PHSD transport approach in a box with periodic boundary conditions in Ref.~\cite{Ozvenchuk13:kubo} as well as in the latest study with 
the DQPM model in Refs.~\cite{Moreau:2019vhw,Soloveva:2019xph}. 
Another way to calculate transport coefficients (explored also in~\cite{Moreau:2019vhw,Soloveva:2019xph}) is to use the relaxation--time approximation (RTA)
\cite{Hosoya:RTA,Chakraborty11,Kapusta,SGavin}.

The shear viscosity based on the RTA (cf.~\cite{Sasaki:2008fg}) reads as:
\begin{align}
\eta^{\text{RTA}}(T,\mu_q)  = \frac{1}{15T} \sum_{i=q,\bar{q},g} \int \frac{d^3p}{(2\pi)^3} \frac{\mathbf{p}^4}{E_i^2}     \tau_i(\mathbf{p},T,\mu) 
\  d_i  (1 \pm f_i) f_i , \label{eta_on} 
\end{align}
where  $d_q = 2N_c = 6$ and $d_g = 2(N_c^2-1) = 16$ are degeneracy factors for spin and color in case of quarks and  gluons , $\tau_i$ are relaxation times. 
Equation~(\ref{eta_on}) includes the Bose enhancement and Pauli-blocking factors, respectively. The~pole energy is $E_i^2 = p^2 + M_i^2$, where $M_i$ is the pole mass given in the DQPM. The~notation $\sum_{j=q,\bar{q},g}$ includes the contribution from all possible partons which in our case are the gluons and the (anti-)quarks of three different flavors ($u,d,s$).

We consider two cases for the relaxation time for quarks and gluons:
(1) $\tau_i(\mathbf{p},T,\mu) =1/ \Gamma_i(\mathbf{p},T,\mu)$
and (2) $\tau_i(T,\mu) =1/ 2\gamma_i(T,\mu)$, 
where $\Gamma_i(\mathbf{p},T,\mu)$ is the parton interaction rate, calculated microscopically from the collision integral using the differential cross sections 
for parton scattering, while $\gamma_i(T,\mu)$ is the width parameter in the parton propagator (\ref{eq:rho}). 

\begin{figure}[H]
\begin{minipage}[h]{0.5\linewidth}
\center{\includegraphics[width=1\linewidth]{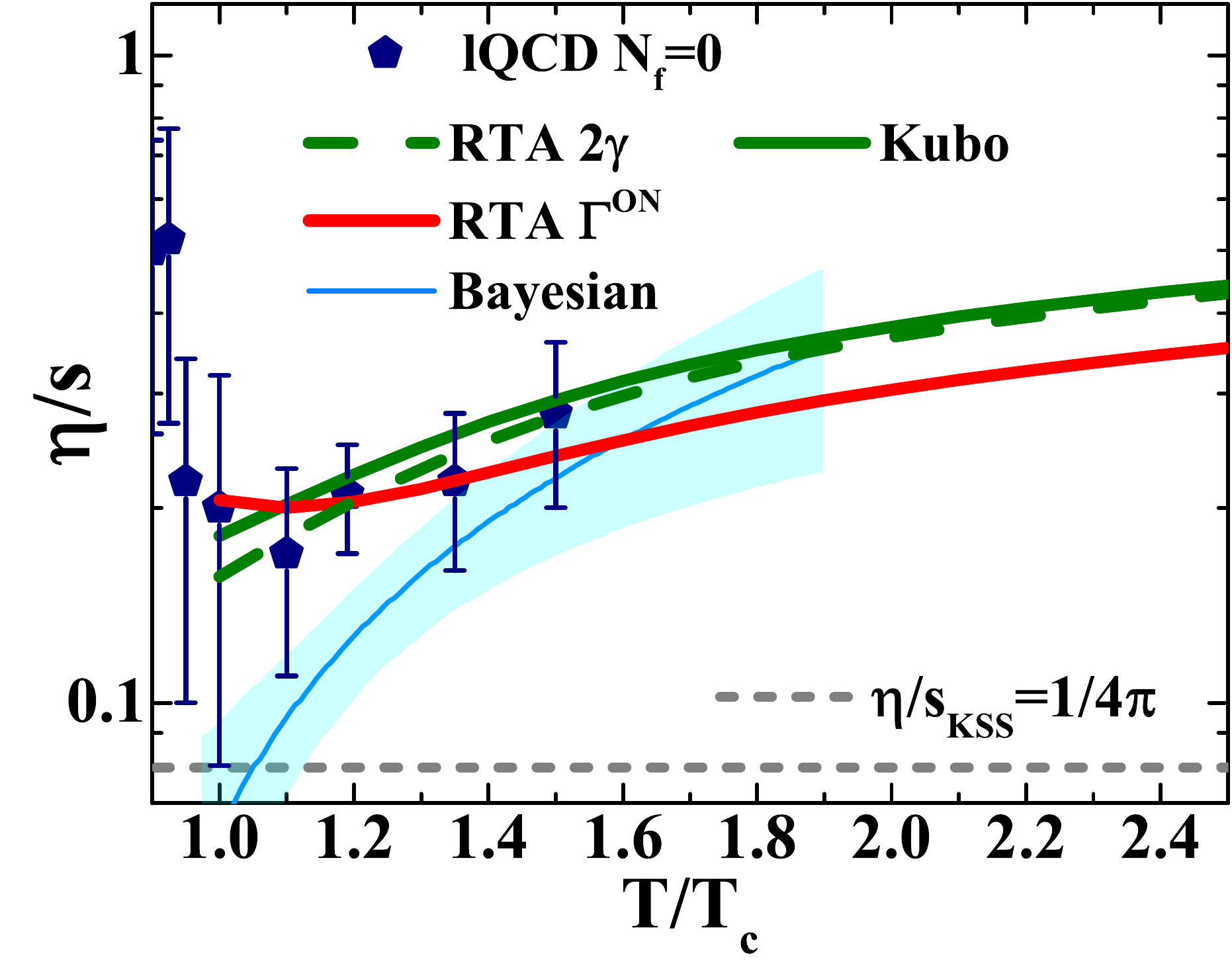} \\ (\textbf{a})}
\end{minipage}
\hfill
\begin{minipage}[h]{0.5\linewidth}
\center{\includegraphics[width=1\linewidth]{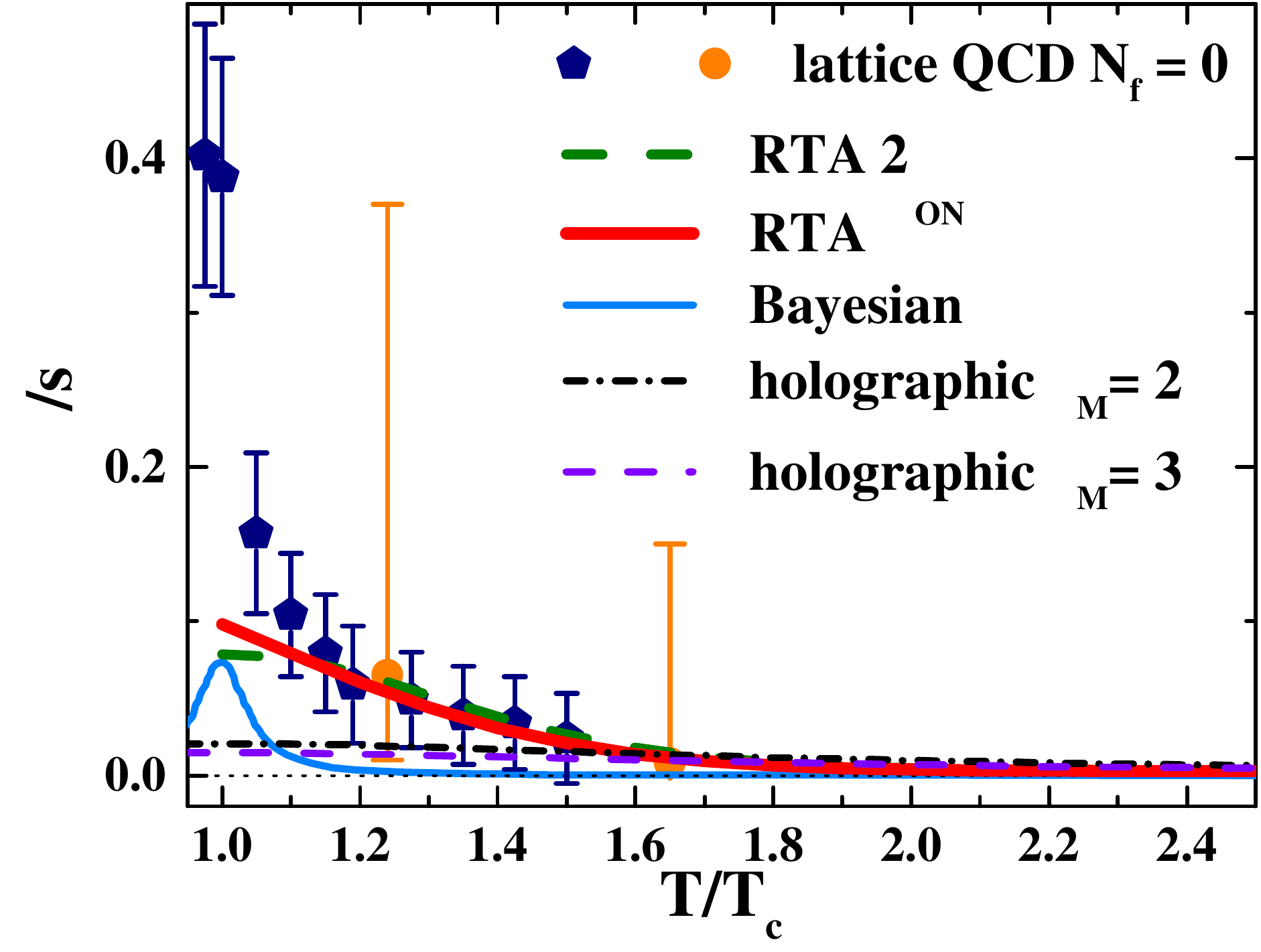} \\ (\textbf{b})}
\end{minipage}
\caption{Left (\textbf{a}): the ratio of shear viscosity to entropy density as a function of 
the scaled temperature $T/T_c$ for $\mu_B = 0$ calculated using the Kubo formalism 
(green solid line) and the RTA approach with the interaction rate $\Gamma^{\rm{on}}$ 
(red solid line) and the DQPM width ${2\gamma}$ (dashed green line).
 The dashed gray line demonstrates the Kovtun--Son--Starinets bound~\cite{Kovtun:2004de} $(\eta/s)_{\rm{KSS}} = 1/(4\pi)$, and~the symbols show lQCD data for pure SU(3) gauge theory taken from Ref.~\cite{Astrakhantsev:2017} (pentagons). The~solid blue line shows the results from a Bayesian analysis of experimental heavy-ion data from Ref.~\cite{Bernhard:2016tnd}.
Right (\textbf{b}): the ratio of the bulk viscosity to entropy density $\zeta/s$ as a function of  the scaled temperature $T/T_c$ for $\mu_B=0$ calculated using the RTA approach with 
the on-shell interaction rate $\Gamma^{\rm{on}}$ (red solid line)
and the DQPM width ${2\gamma}$ (dashed green line). The~symbols correspond to the lQCD data for pure SU(3) gauge theory taken from Refs.~\cite{Astrakhantsev:2018oue} (pentagons) and~\cite{Meyer} (circles). The~solid blue line shows the results from a Bayesian analysis of experimental heavy-ion data from Ref.~\cite{Bernhard:2016tnd}.
The dot-dashed and dashed lines correspond to the results from the non-conformal holographic model for $\phi_M=2$ and 3, correspondingly, from~Ref.~\cite{Maximilian}.}
\label{fig_eta_zeta}
\end{figure}

In the left part (a) of Figure~\ref{fig_eta_zeta}, we show
the ratio of the shear viscosity to entropy density as a function of the scaled temperature 
$T/T_c$ for $\mu_B = 0$ calculated using the Kubo formalism (green solid line) and RTA approach
with the interaction rate $\Gamma^{\rm{on}}$ (red solid line) and the DQPM width ${2\gamma}$
(dashed green line).
The RTA approximation (\ref{eta_on}) of the shear viscosity with the DQPM width
${2\gamma}$ and with the interaction rate $\Gamma^{\rm{on}}$ are quite close to each other
at $\mu_B=0$ and also very close to the one from the Kubo formalism~\cite{Moreau:2019vhw}  indicating that the quasiparticle limit ($\gamma \ll M$) holds in the~DQPM.

The ratio $\eta/s$ increases with an increase of the scaled temperature. 
The actual values for the ratio $\eta/s$  are in a good agreement with the gluodynamic
lattice QCD calculations at $\mu_B=0$ from Ref.~\cite{Astrakhantsev:2017}. 
Moreover, our DQPM results are in qualitative agreement with the results 
from a Bayesian analysis of experimental heavy-ion data from Ref.~\cite{Bernhard:2016tnd}.
We mention that the DQPM result differs from the recent calculations for the 
shear viscosity at $\mu_B=0$ in the quasiparticle model in Ref.~\cite{Mykhaylova:2019}
where the width of quasiparticles is not considered which leads to a high 
value for the $\eta/s$ ratio. This shows the sensitivity of this ratio to the modelling of
partonic interactions and the properties of partons in the hot QGP medium.
We remind also that in Refs.~\cite{Moreau:2019vhw,Soloveva:2019xph,Soloveva:2019doq} 
we find that the ratio $\eta/s$ shows a very weak dependence on $\mu_B$
and has a similar behavior as a function of temperature for all $\mu_B \le 400$ MeV.

The expression for the bulk viscosity of the partonic phase derived within the RTA 
 reads (following  Ref.~\cite{Kapusta})
\begin{align}
\zeta^{\text{RTA}}(T,\mu)= \frac{1}{9T} \sum_{i=q,\bar{q},g}\int \frac{d^3p}{(2\pi)^3}
 \tau_i(\mathbf{p},T,\mu) 
\ \frac{ d_i  (1 \pm f_i) f_i  }{E_i^2}\left(\mathbf{p}^2-3c_s^2\left(E_i^2-T^2\frac{dm_i^2}{dT^2}\right)\right)^2 , 
  \label{zeta_on} \end{align}
where $c_s^2$ is the speed of sound squared, and $\frac{dm_i^2}{dT^2}$ is the DQPM parton mass derivative which becomes large close to the critical temperature $T_c$.

On the right side (b) of Figure~\ref{fig_eta_zeta}, we show the ratio of the bulk viscosity to
entropy density $\zeta/s$ as a function of  the scaled temperature $T/T_c$ for $\mu_B=0$
calculated using the RTA approach with the interaction rate $\Gamma^{\rm{on}}$ 
(red solid line) and the DQPM width ${2\gamma}$ (dashed green line).
The symbols correspond to the lQCD data for pure SU(3) gauge theory taken from 
Refs.~\cite{Astrakhantsev:2018oue} (pentagons) and~\cite{Meyer} (circles). 
The solid blue line shows the results from a Bayesian analysis of experimental 
heavy-ion data from Ref.~\cite{Bernhard:2016tnd}. 
The dot-dashed and dashed lines correspond to the results from the non-conformal holographic
model~\cite{Maximilian} for $\phi_M=2$ and 3, correspondingly, where $\phi_M$
is the model parameter which characterizes the non-conformal features of the model.
We find that the DQPM result for $\zeta/s$ is in very good agreement with the lattice QCD
results and shows a rise closer to
$T_C$ contrary to the holographic results, which show practically a constant behavior 
independent of  model parameters. This rise is attributed to the increase of the partonic
mass closer to  $T_C$ as shown in Figure~\ref{fig1}, thus the mass derivative term
in Equation~(\ref{zeta_on}) also grows. The~Bayesian result also shows a peak near $T_C$;
however, the~ratio drops to zero while lQCD data indicate the positive $\zeta/s$
as found also in the DQPM.
The~$\mu_B$ dependence of $\zeta/s$ has been investigated within the DQPM in Refs.
~\cite{Moreau:2019vhw,Soloveva:2019xph,Soloveva:2019doq}, where it has been shown that
it is rather weak for $\mu_B \le 400$ MeV, similar to $\eta/s$.
As follows from hydrodynamical calculations, the~results for the flow harmonic $v_n$
is sensitive to the transport coefficients~\cite{Romatschke,Song:Heinz,Bernhard:2016tnd}.
Thus, there are hopes to observe a $\mu_B$ sensitivity of $v_1, v_2$.

\section{Heavy-Ion~Collisions}

In our recent study~\cite{Moreau:2019vhw}, we have investigated the sensitivity 
of 'bulk' observables such as rapidity and transverse momentum distributions 
of different hadrons produced in heavy-ion collisions from AGS to top RHIC energies 
on the details of the QGP interactions and the properties of partonic degrees-of-freedom.
For that, we have considered the following three cases:

(1) {\bf `PHSD4.0'}: the masses and widths of quarks and gluons depend only on $T$.
The cross sections for partonic interactions depend only on $T$ as evaluated in 
the `box' calculations in Ref.~\cite{Ozvenchuk13} in order to merge the QGP interaction
rates from all possible partonic channels to the total temperature dependent widths 
of the DQPM propagator. This has been used in the PHSD code (v. 4.0 or below) for
extended studies of many hadronic observables in p+A and A+A collisions 
at different energies~\cite{Cassing:2008sv,Cassing:2008nn,Cassing:2009vt,Bratkovskaya:2011wp,Linnyk:2015rco,Konchakovski:2011qa}. 

(2) {\bf `PHSD5.0 - $\mu_B=0$'}: the masses and widths of quarks and gluons depend only 
on $T$; however, the~differential and total partonic cross sections are obtained by calculations of the leading order Feynman 
diagrams employing the effective propagators and couplings $g^2(T/T_c)$ from the DQPM 
at $\mu_B=0$ \cite{Moreau:2019vhw}. Thus, the~cross sections depend explicitly on the
invariant energy of the colliding partons $\sqrt{s}$ and on $T$. This is realized in the PHSD5.0 by
setting $\mu_B=0$, cf.~\cite{Moreau:2019vhw}.

(3) {\bf `PHSD5.0 - $\mu_B$'}: the masses and widths of quarks and gluons depend on $T$
and $\mu_B$ explicitly; the differential and total partonic cross sections are obtained by calculations of the leading order Feynman diagrams from the DQPM and explicitly depend on invariant energy $\sqrt{s}$, temperature $T$ and baryon chemical potential $\mu_B$. 
This is realized in the full version of PHSD5.0, cf.~\cite{Moreau:2019vhw}.

The comparison of the 'bulk' observables for A+A collisions within the three cases of PHSD
in Ref.~\cite{Moreau:2019vhw}
has illuminated that they show a very low sensitivity to the $\mu_B$ dependences of parton
properties (masses and widths) and their interaction cross sections such that the results
from PHSD5.0 with and without $\mu_B$ were very close to each other. 
Only in the case of kaons,  antiprotons $\bar{p}$ and antihyperons $\bar{\Lambda} + \bar{\Sigma}^0$, a~small difference between PHSD4.0 and PHSD5.0 could be seen at 
top SPS and top RHIC energies. A~similar trend has been found for very asymmetric collisions of C+Au: a small sensitivity to the partonic scatterings was found in the kaon and antibaryon rapidity distributions too.
This could be understood as following: at high energies such as top RHIC where the QGP volume
is very large in central collisions, the~$\mu_B$ is very low, while, when
decreasing the bombarding energy---in order to increase $\mu_B$,
the fraction of the QGP is decreasing such that the final observables
are dominated by the hadronic phase, i.e.,~the probability for 
the hadrons created at the QGP hadronization to re-scatter, decay, or be absorbed 
in hadronic matter increases strongly; accordingly, the sensitivity to the properties of the QGP is washed~out.

\subsection{Asymmetric~Systems}
In Ref.~\cite{Moreau:2019vhw}, we have investigated the sensitivity to $\mu_B$  of 
the `bulk' observables in asymmetric heavy-ion collisions for C+Au.
The spectra for  C+Au indicated that they show a slightly larger sensitivity to $\mu_B$ for antiprotons and strange hadrons, kaons and antihyperons than for pions and protons.  
Here, we present the results for the asymmetric Cu+Au~collisions. 

\begin{figure}[H]\phantom{a}
\begin{minipage}[H]{0.7\linewidth}
\includegraphics[width=.8\linewidth]{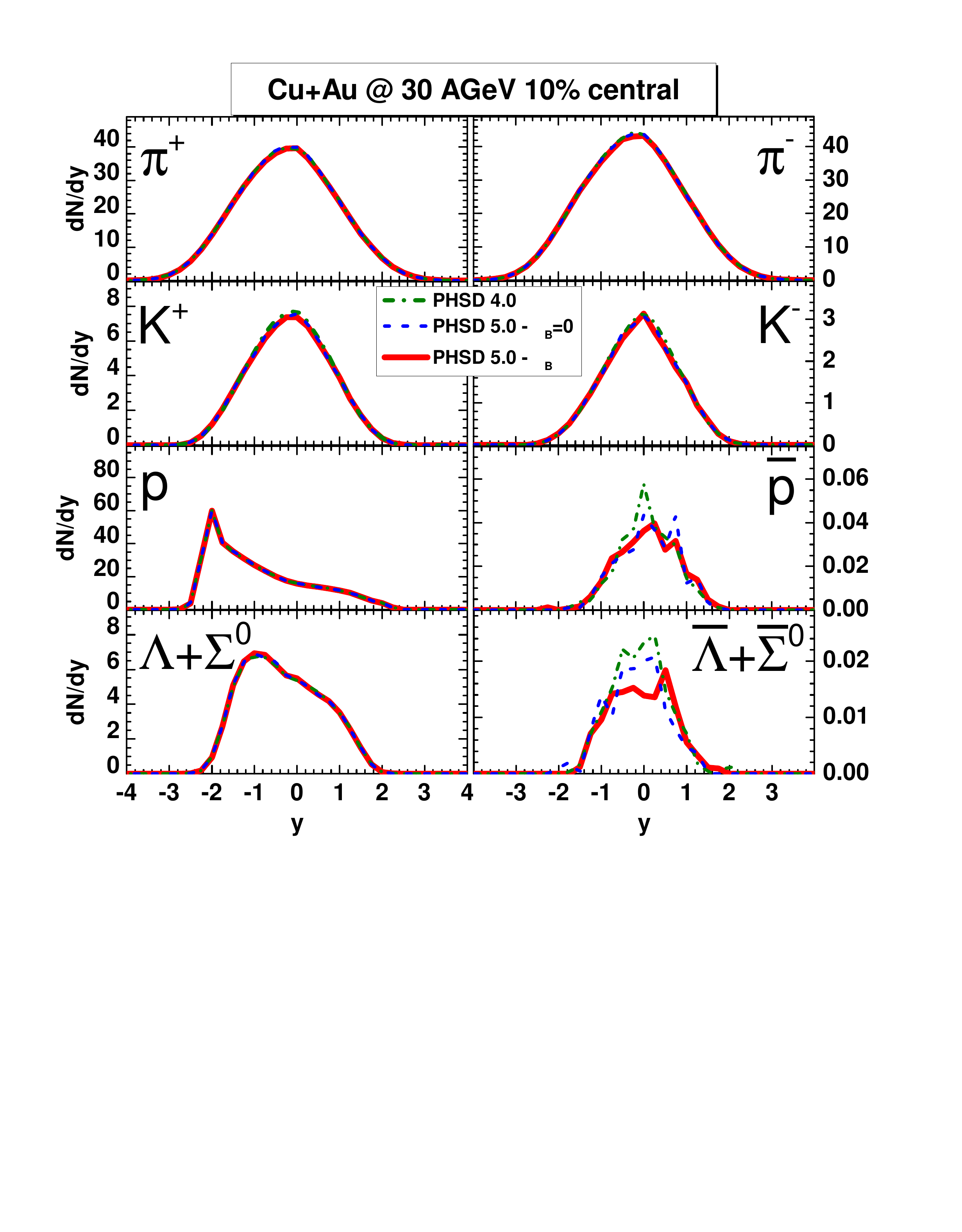} 
\end{minipage}\phantom{a}\hspace*{-1.5cm}
\begin{minipage}[H]{0.7\linewidth}
\includegraphics[width=0.5\linewidth]{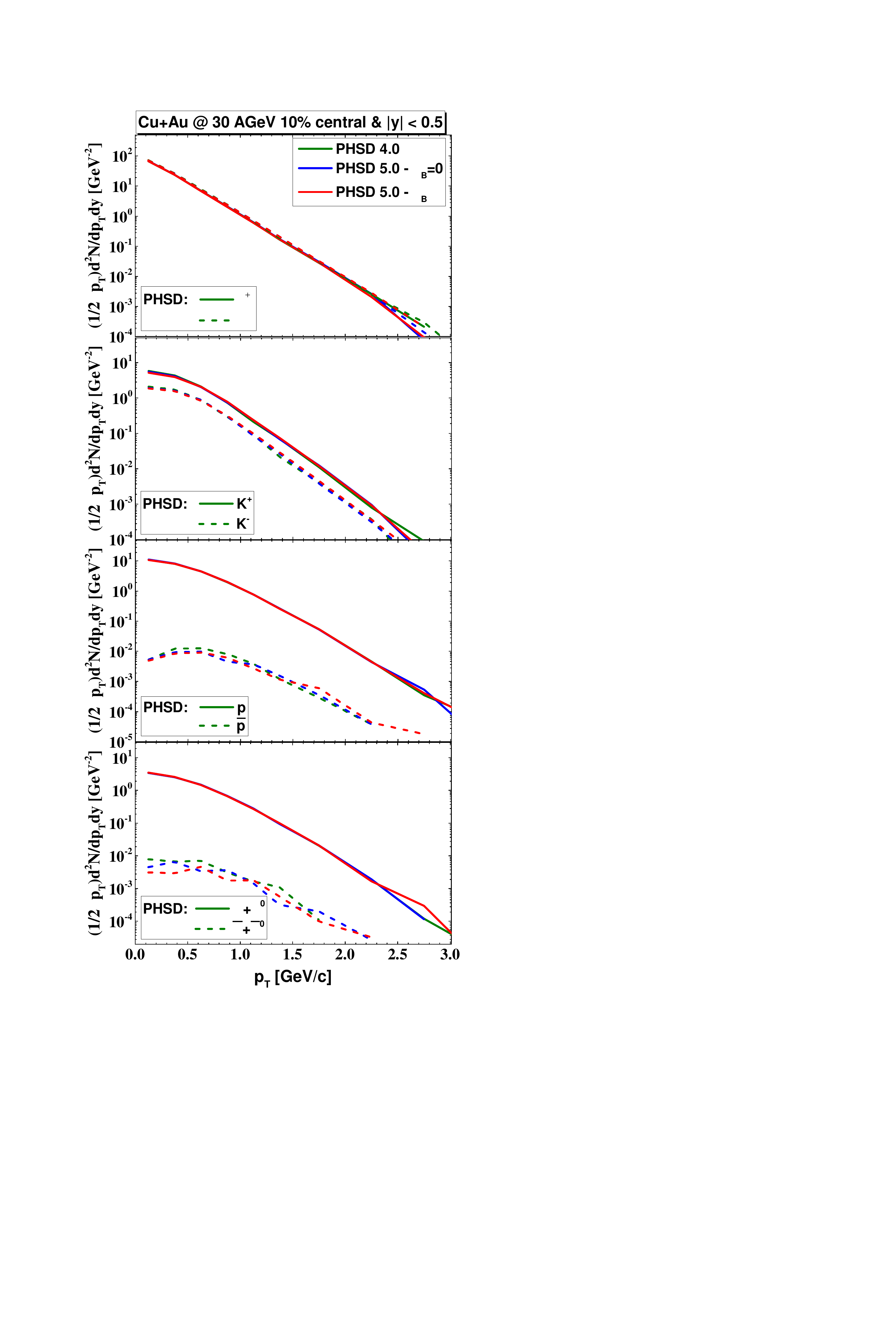} 
\end{minipage}
\caption{The rapidity distributions (left plot) and $p_T$-spectra at 
midrapidity ($|y| < $ 0.5) (right plot) 
for $\pi^\pm, K^\pm, p, \bar p, \Lambda+\Sigma^0, \bar\Lambda + \bar\Sigma^0$
for 10\% central Cu+Au collisions at 30 A GeV for PHSD4.0 (green dot-dashed lines), PHSD5.0 with partonic cross sections and parton masses calculated for $\mu_B$ = 0 (blue dashed lines) and with cross sections and parton masses evaluated at the actual chemical potential $\mu_B$ in each individual space-time cell (red lines).}
\label{CuAu30}
\end{figure}

\begin{figure}[H]
\begin{minipage}[H]{0.7\linewidth}\phantom{a}
\center{\includegraphics[width=.78\linewidth]{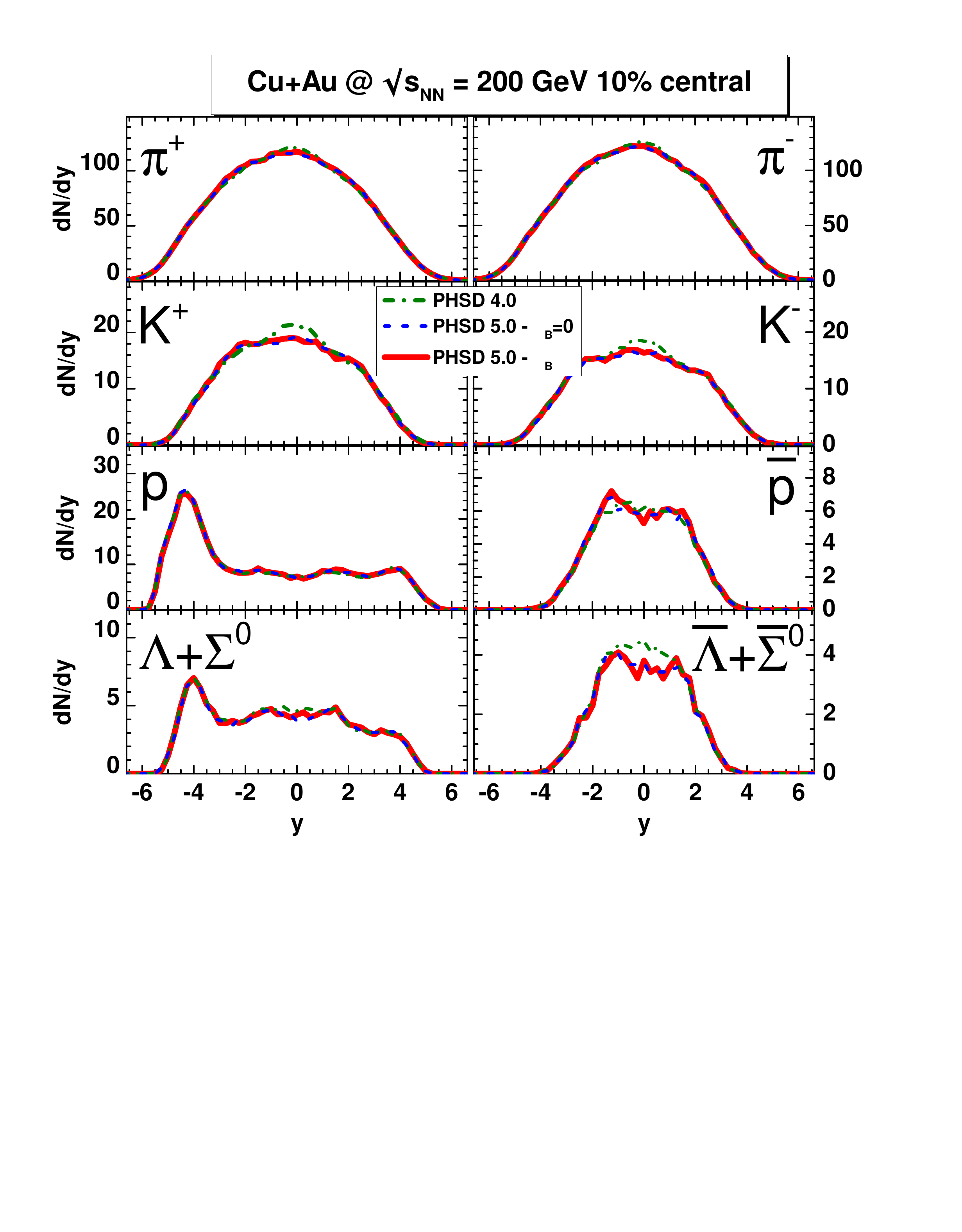}}
\end{minipage}\phantom{a}\hspace*{-4cm}
\begin{minipage}[H]{0.7\linewidth}
\center{\includegraphics[width=.5\linewidth]{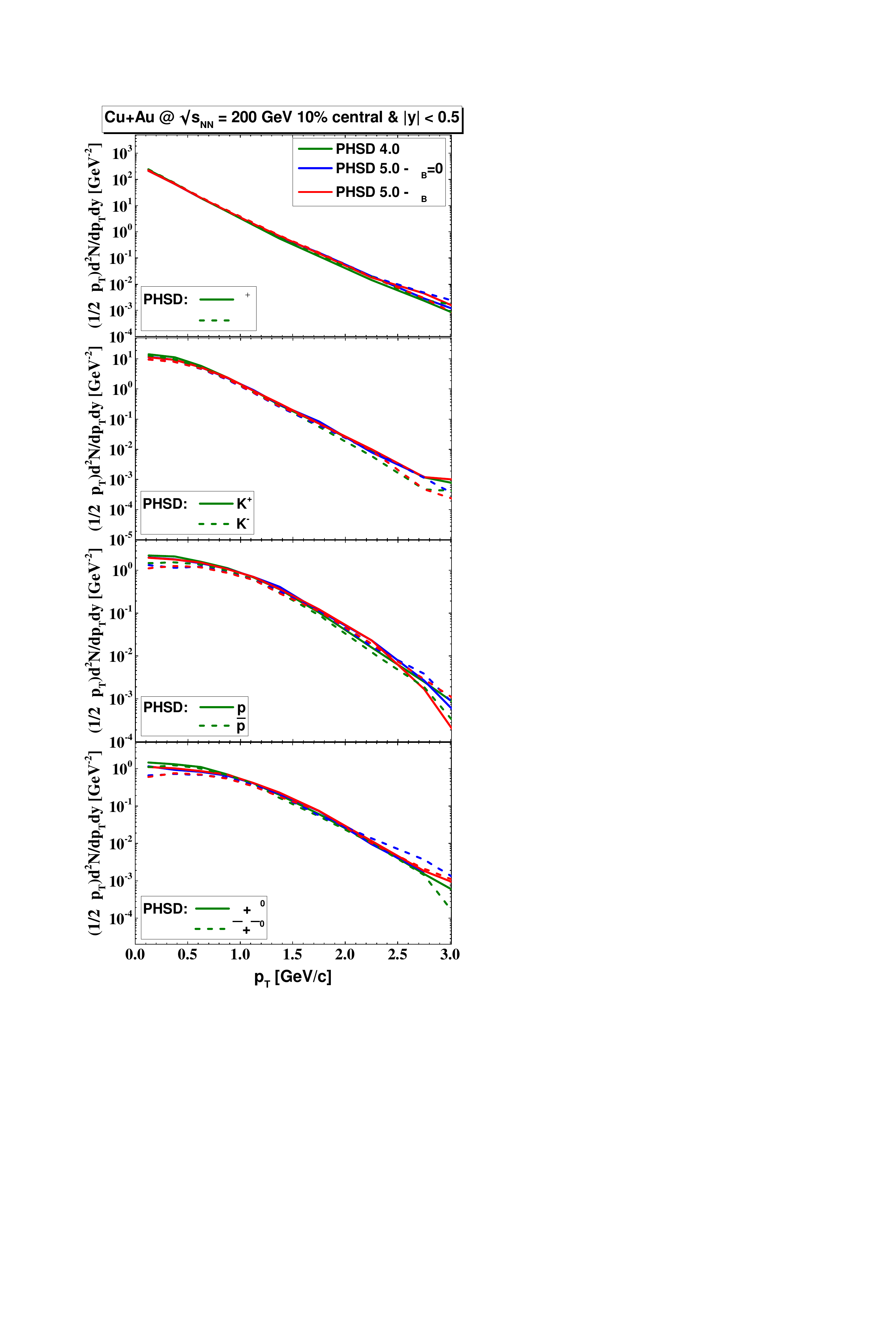}}
\end{minipage}
\caption{The rapidity distributions (left plot) and $p_T$-spectra at 
midrapidity ($|y| < $ 0.5) (right plot) 
for $\pi^\pm, K^\pm, p, \bar p, \Lambda+\Sigma^0, \bar\Lambda + \bar\Sigma^0$
for 10\% central Cu+Au collisions at $\sqrt{s_{NN}}=200$ GeV for PHSD4.0 (green dot-dashed lines), PHSD5.0 with partonic cross sections and parton masses calculated for $\mu_B$ = 0 (blue dashed lines) and with cross sections and parton masses evaluated at the actual chemical potential $\mu_B$ in each individual space-time cell (red lines).}
\label{CuAu200}
\end{figure}

In Figures~\ref{CuAu30} and \ref{CuAu200}, we show the rapidity distributions (left plot) and 
$p_T$-spectra at midrapidity ($|y| < $ 0.5) (right plot) 
for $\pi^\pm, K^\pm, p, \bar p, \Lambda+\Sigma^0, \bar\Lambda + \bar\Sigma^0$
for  10\% central Cu+Au collisions at 30 AGeV and  $\sqrt{s_{NN}}=200$ GeV for three cases:
(1) PHSD4.0 (green dot-dashed lines), (2) PHSD5.0 with partonic cross sections and parton masses/widths calculated for $\mu_B$ = 0 (blue dashed lines) and (3) with cross sections and parton masses/widths evaluated at the actual chemical potential $\mu_B$ in each individual space-time cell (red lines).
Similar to C+Au collisions, we find for Cu+Au collisions a small difference in the rapidity
distributions of antiprotons and in the strangeness sector - in kaon and especially 
in  $\bar\Lambda + \bar\Sigma^0$ $y$-distributions. 
Similar statements hold for the $p_T$ spectra which show a slightly different slope at low
and high momenta of anti-strange baryons.
This suggests that the strange degree-of-freedom might be experimentally explored in 
asymmetric systems to obtain additional information on the partonic~interactions.

\unskip

\subsection{Directed~Flow}


Now, we test the traces of $\mu_B$ dependences of the QGP interaction cross sections
in collective observables such as directed flow $v_1$ considering again three cases
of the PHSD as discussed~above.

\begin{figure}[H]
\centering
\includegraphics[width=10cm]{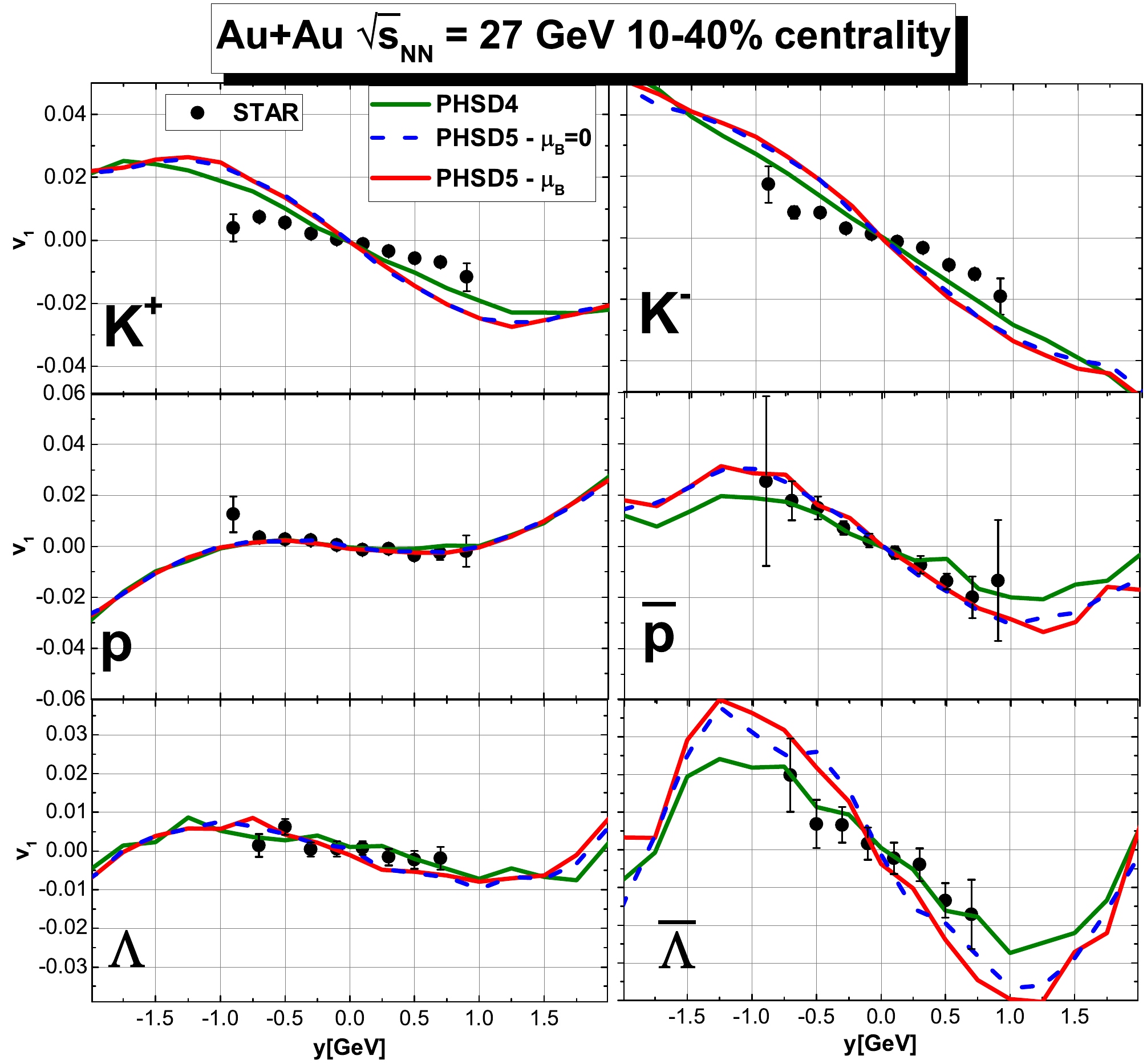} 
\caption{Directed flow of identified hadrons as a function of rapidity at $\sqrt{s_{NN}}$ = 27 GeV for PHSD4.0 (green lines), PHSD5.0 with partonic cross sections and parton masses calculated for $\mu_B$ = 0 (blue dashed lines) and with cross sections and parton masses evaluated at the actual chemical potential $\mu_B$ in each individual space-time cell (red lines) in comparison to the experimental data of the STAR Collaboration~\cite{STARBESv1}.}
\label{v1y27GeV}
\end{figure}

Figure~\ref{v1y27GeV} depicts the directed flow $v_1$ of identified hadrons
($K^\pm, p, \bar p, \Lambda+\Sigma^0, \bar\Lambda + \bar\Sigma^0$) versus
rapidity for $\sqrt{s_{NN}}$ = 27 GeV. 
One can see a good agreement between PHSD results and experimental data from STAR collaboration~\cite{STARBESv1}.
However, the~different versions of PHSD for the $v_1$ coefficient show a quite
similar behavior; only antihyperons show a slightly different flow.
This supports again the finding that strangeness, and~in particular anti-strange
hyperons, are the most sensitive probes for the QGP~properties.

\subsection{Elliptic~Flow}
As follows from the hydrodynamic simulations~\cite{Romatschke,Song:Heinz} and from
the Bayesian analysis~\cite{Bernhard:2016tnd}, an~elliptic flow $v_2$ is sensitive 
to the transport properties of the QGP characterized by transport coefficients such
as shear $\eta$ and bulk $\zeta$ viscosities.
In this section, we present the results for the elliptic flow of charged hadrons 
from HIC within the PHSD5.0 with and without $\mu_B$ dependences and compare the results 
with PHSD4.0, again.
 
\begin{figure}[H]
\begin{minipage}[h]{0.32\linewidth}
\center{\includegraphics[width=1\linewidth]{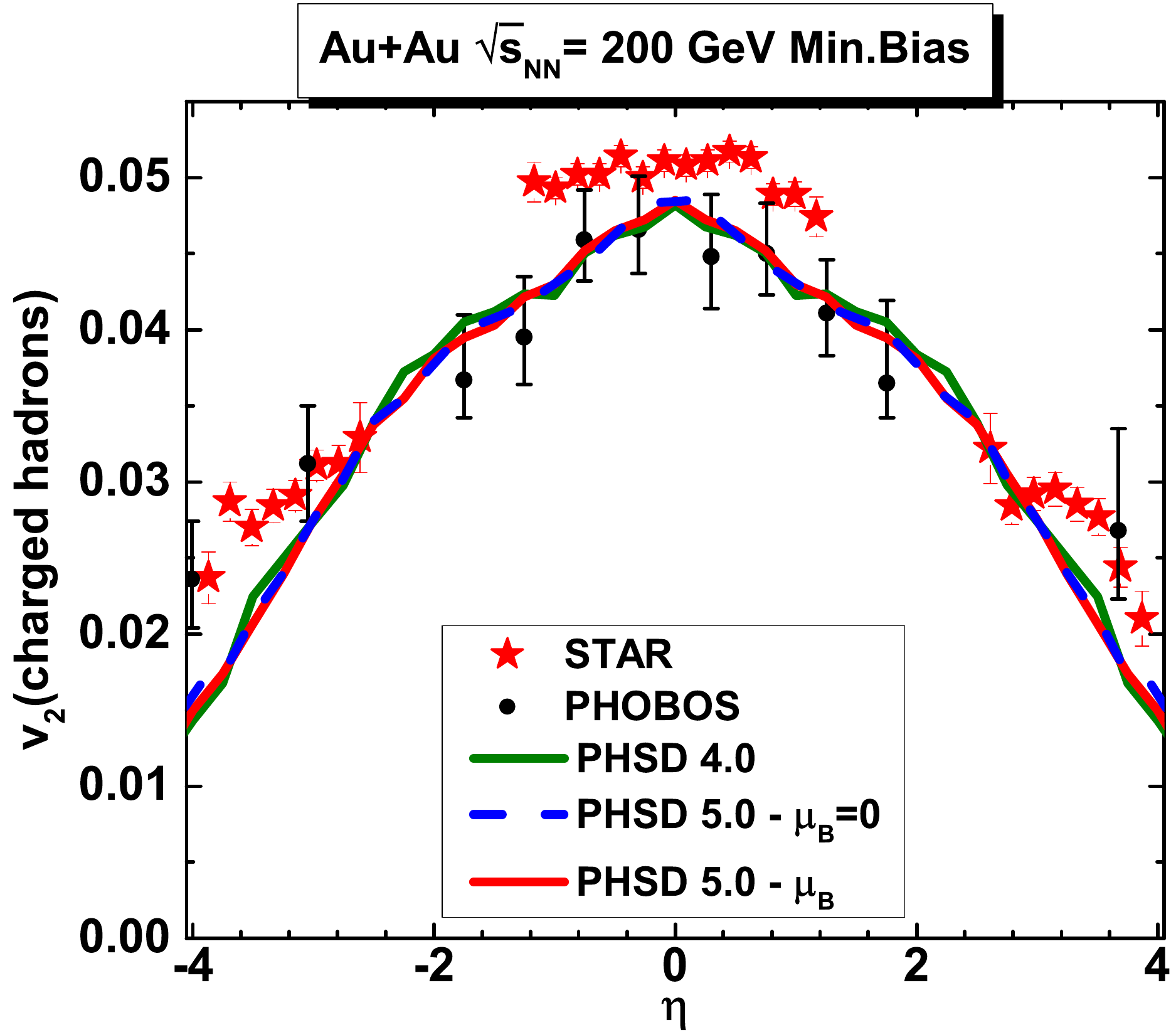} \\ (\textbf{a})}
\end{minipage}
\hfill
\begin{minipage}[h]{0.32\linewidth}
\center{\includegraphics[width=1\linewidth]{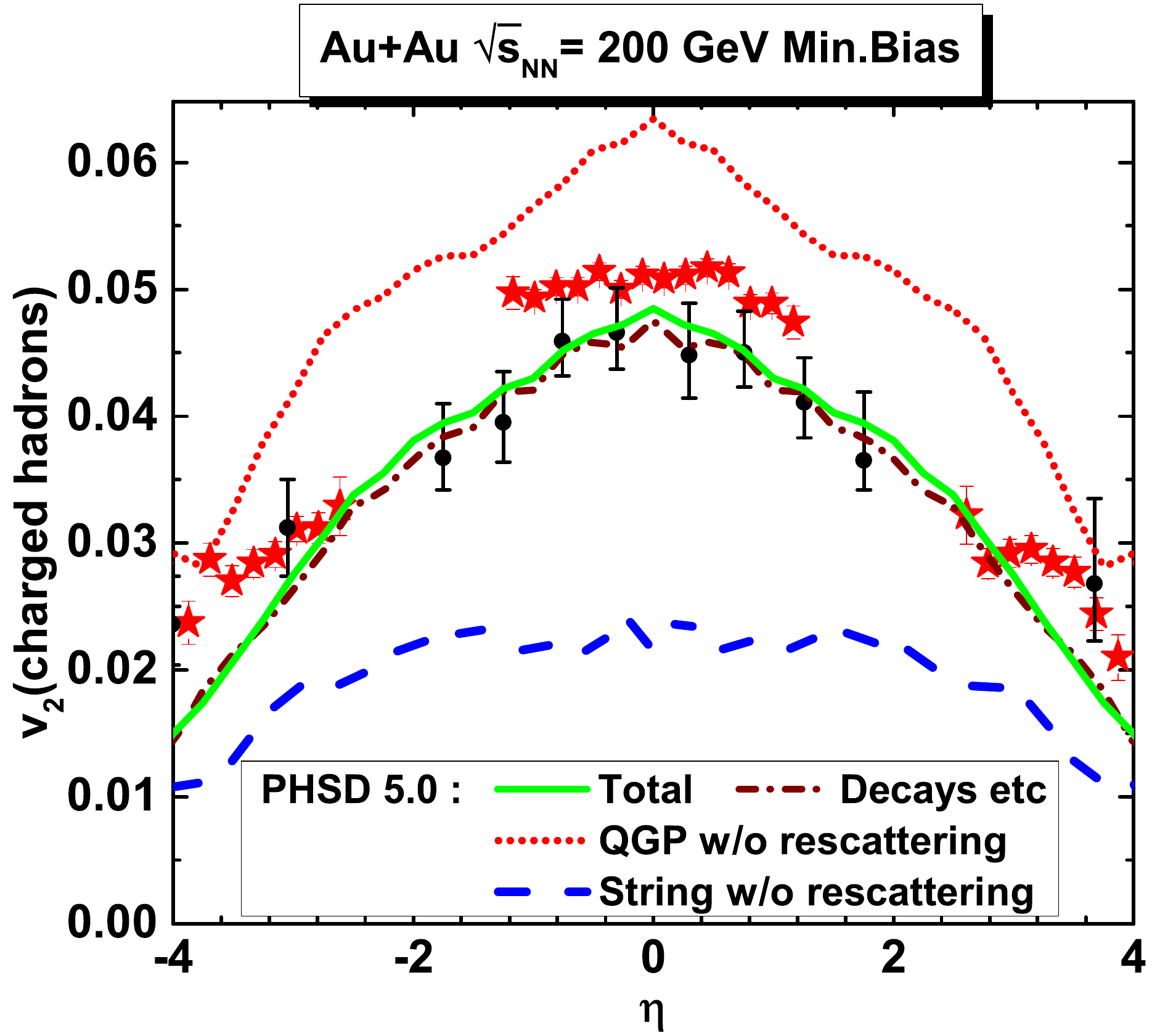} \\ (\textbf{b})}
\end{minipage}
\begin{minipage}[h]{0.32\linewidth}
\center{\includegraphics[width=1\linewidth]{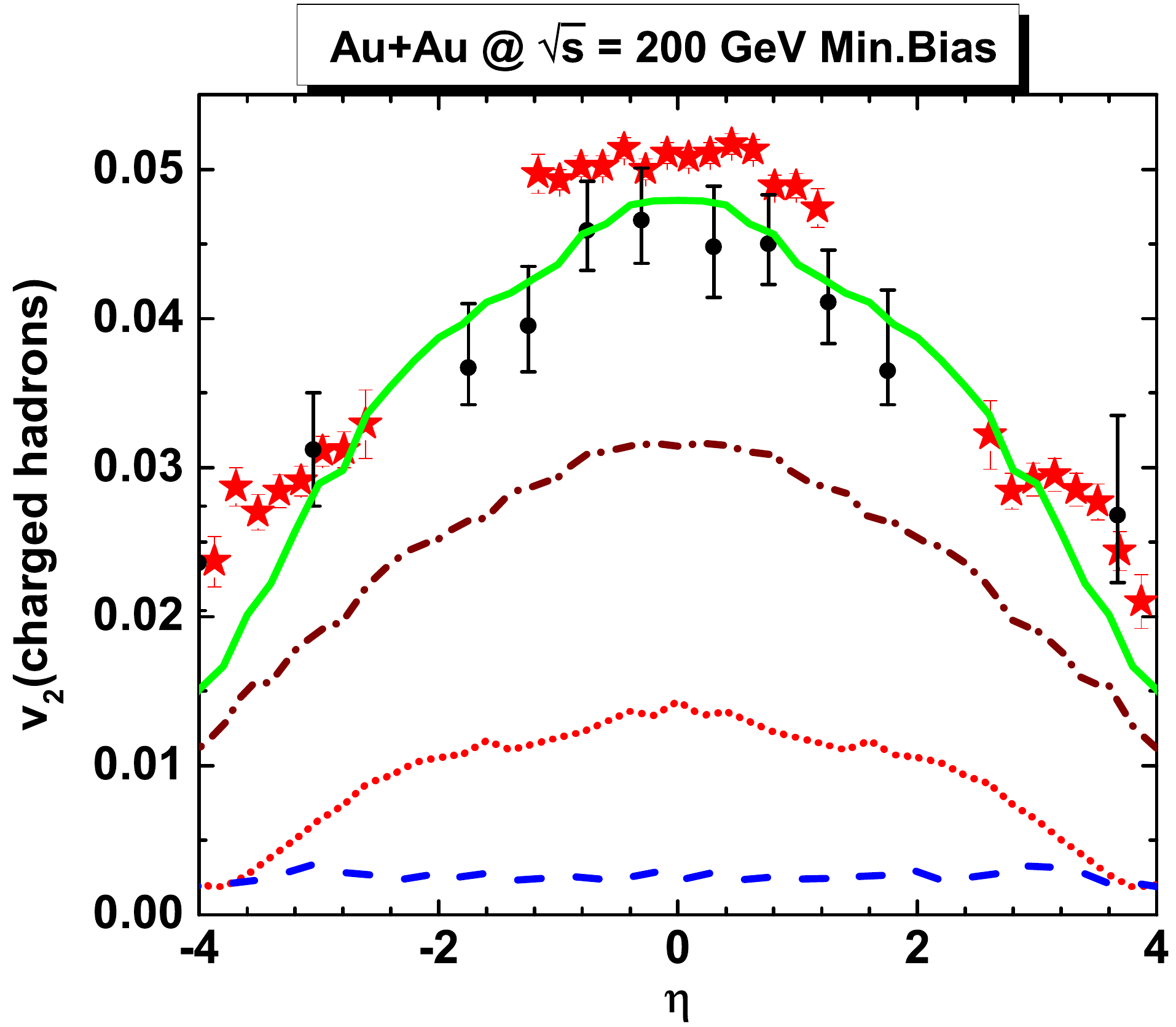} \\ (\textbf{c})}
\end{minipage}
\caption{Left (\textbf{a}): elliptic flow of charged hadrons as function of pseudo-rapidity $\eta$  
for minimum bias Au+Au collisions at $\sqrt{s_{NN}}$=200 GeV for PHSD4.0 (green lines), PHSD5.0 with partonic cross sections and parton masses calculated for $\mu_B$ = 0 (blue dashed lines), and~with the actual $\mu_B$ (red lines) in comparison to the experimental data from  STAR~\cite{STARv2_200} (solid starts)
and PHOBOS~\cite{PHOBOS} (solid dots).
Middle (\textbf{b}): individual contributions to $v_2$ without their relative weights to the 
total $v_2$, which are indicated by a green solid line for PHSD5.0 with $\mu_B$: 
the red dotted line corresponds to the final  hadrons coming from the QGP 
without rescattering in the hadronic phase, the~blue dashed line indicates the $v_2$  
of hadrons coming from strings while the brown dot-dashed line shows the $v_2$ 
of hadrons coming from mesonic and baryonic resonance~decays.
Right (\textbf{c}): individual contributions to $v_2$ including their relative weights to the 
total $v_2$.\\}
\label{ris:v2eta}
\end{figure}

\begin{figure}[H]
\begin{minipage}[h]{0.32\linewidth}
\center{\includegraphics[width=1\linewidth]{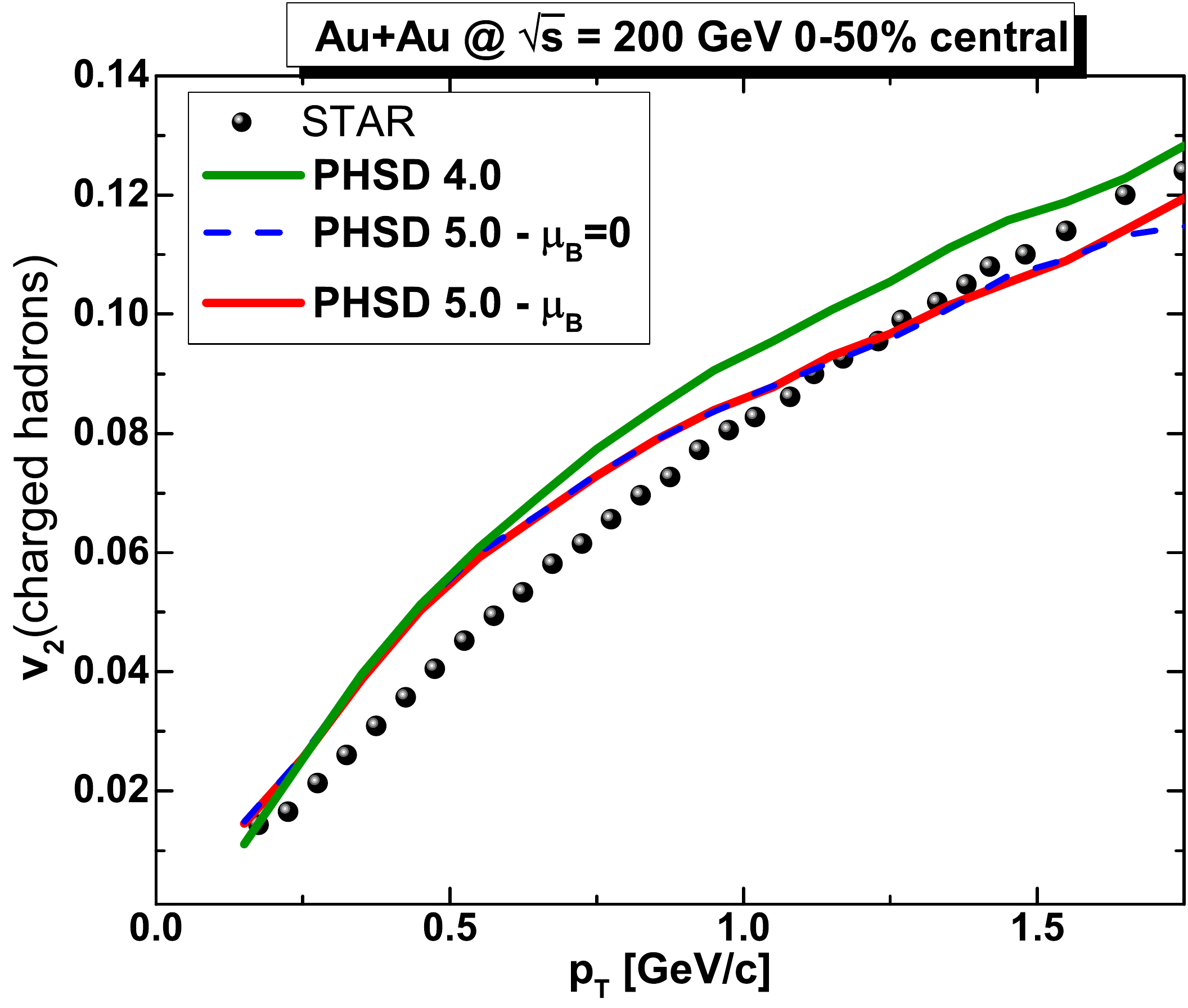} \\ (\textbf{a})}
\end{minipage}
\hfill
\begin{minipage}[h]{0.32\linewidth}
\center{\includegraphics[width=1\linewidth]{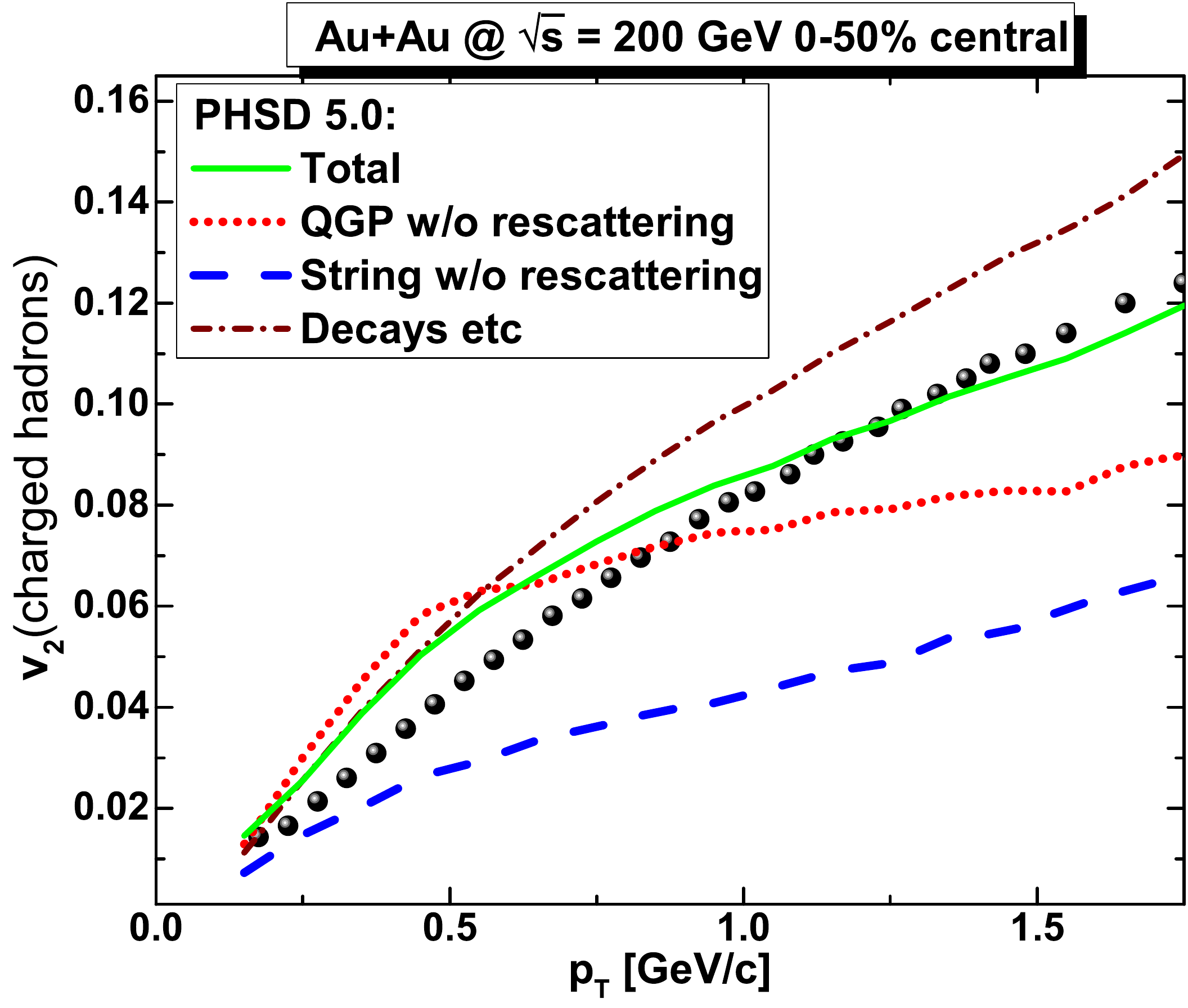} \\ (\textbf{b})}
\end{minipage}
\begin{minipage}[h]{0.32\linewidth}
\center{\includegraphics[width=1\linewidth]{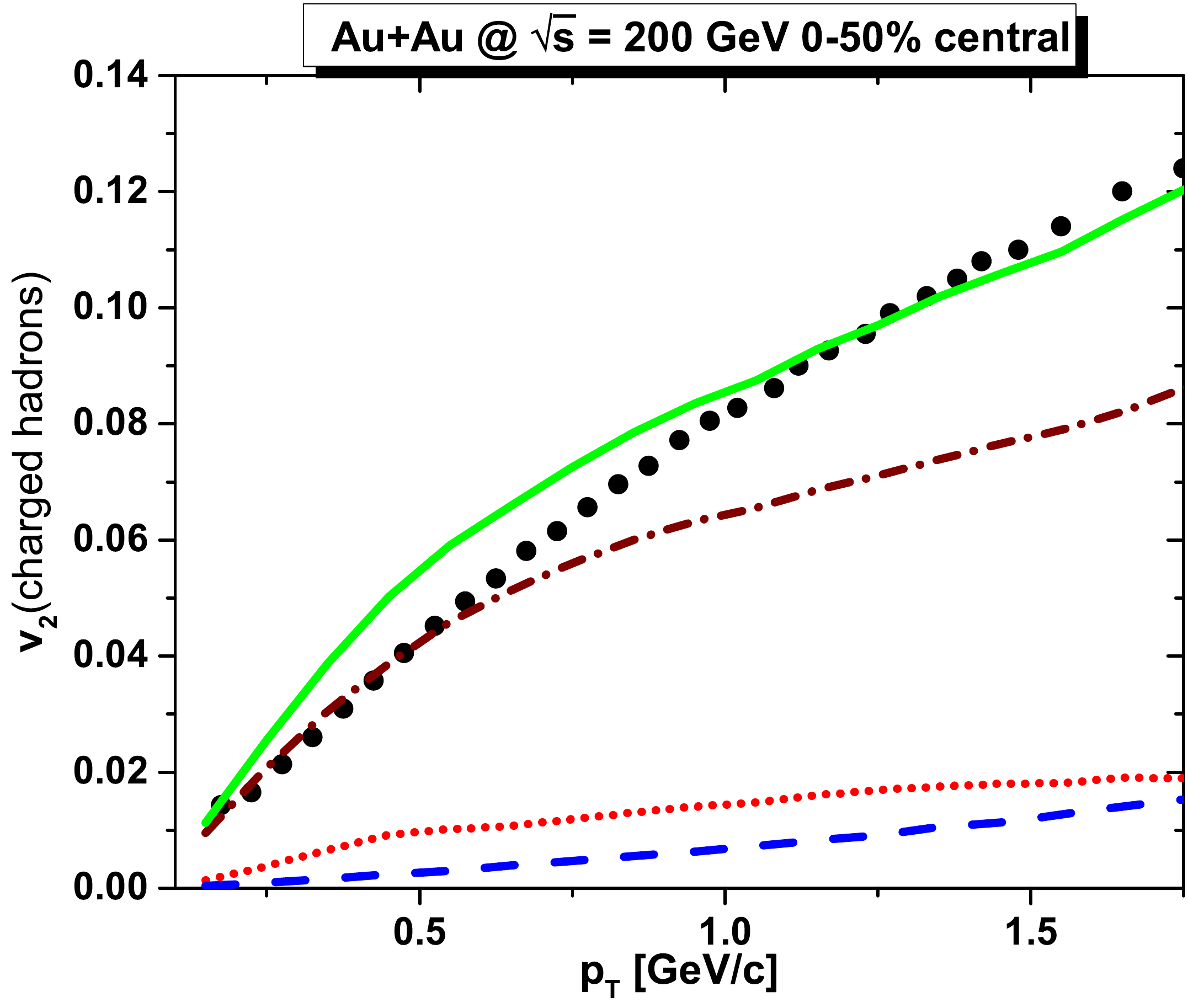} \\ (\textbf{c})}
\end{minipage}
\caption{Elliptic flow of charged hadrons as a function of $p_T$ for 0--50\% central Au+Au collisions at $\sqrt{s_{NN}}$ = 200 GeV.
The line description is similar to Figure~\ref{ris:v2eta}.}
\label{ris:v2pt}
\end{figure}

The left plots `(a)' in Figures~\ref{ris:v2eta} and \ref{ris:v2pt} display the actual results 
for charged hadron elliptic flow  as a function of pseudo-rapidity $\eta$ (Figure \ref{ris:v2eta}) and of transverse momentum $p_T$ (Figure \ref{ris:v2pt})
for  minimum bias Au+Au collisions at $\sqrt{s_{NN}}$ = 200 GeV for PHSD4.0 (green lines), PHSD5.0 with partonic cross sections and parton masses calculated for $\mu_B$ = 0 (blue dashed lines), and~with cross sections and parton masses evaluated at the actual chemical potential $\mu_B$ in each individual space-time cell (red lines) in comparison to the experimental data from  the  STAR collaboration~\cite{STARv2_200} (solid stars)
and PHOBOS~\cite{PHOBOS} (solid dots).
One can see the difference for $v_2(p_T)$ in case of charged hadrons for high $p_T> 0.5$ GeV between PHSD4.0 and~PHSD5.0.  

\unskip

The channel composition of $v_2$ for  PHSD5.0---with cross sections and parton masses
evaluated at the actual chemical potential $\mu_B$ in each individual space-time cell---is shown in  the middle plots `(b)' of Figures~\ref{ris:v2eta} and \ref{ris:v2pt}.  
We sorted the particles according to their production channels into three parts:
the red dotted line corresponds to the final  hadrons coming from the QGP 
without rescattering in the hadronic phase, the~blue dashed line indicates the $v_2$  
of hadrons coming from strings (without further rescattering) while the brown 
dot-dashed line shows the $v_2$ of hadrons coming from mesonic and baryonic 
resonance decays. 
One can see a large difference between the averaged elliptic flow for the different channels:
the $v_2$ of hadrons from string decay is the lowest since string production occurs dominantly
at the initial phase of the heavy-ion collision;
the $v_2$ of hadrons from the QGP is the largest versus $\eta$ as follows from the middle Part `(b)'
of Figure~\ref{ris:v2eta}. However, this is mainly due to the low $p_T$ hadrons which
give a larger contribution to $v_2(\eta)$---cf. the middle part `(b)' of 
Figure~\ref{ris:v2pt}. Here, the high $p_T$ hadrons from the QGP show a lower $v_2$ 
than those coming from strings or resonance~decays.

The right parts `(c)' of Figures~\ref{ris:v2eta} and \ref{ris:v2pt} present 
the individual contributions to $v_2$ including their relative weights to the total $v_2$. It shows that the properly weighted channel decomposition of $v_2$ looks rather different---the contribution of the hadrons from the QGP is now small since most of them rescatter in the hadronic phase, i.e.,~the relative fraction of hadrons directly coming from
QGP hadronization is very small. The~total $v_2$ is dominated by the hadrons
coming from the decay of resonances. The~fraction of hadrons from string decays
is very small due to the fact that strings are formed mainly in the beginning of 
collisions, and~a very small fraction of hadrons can survive directly. 
Thus, the~information in $v_2$ about the QGP properties is washed out to a large extent
by final hadronic~interactions.

\begin{figure}[H]
\centering
\includegraphics[width=10cm]{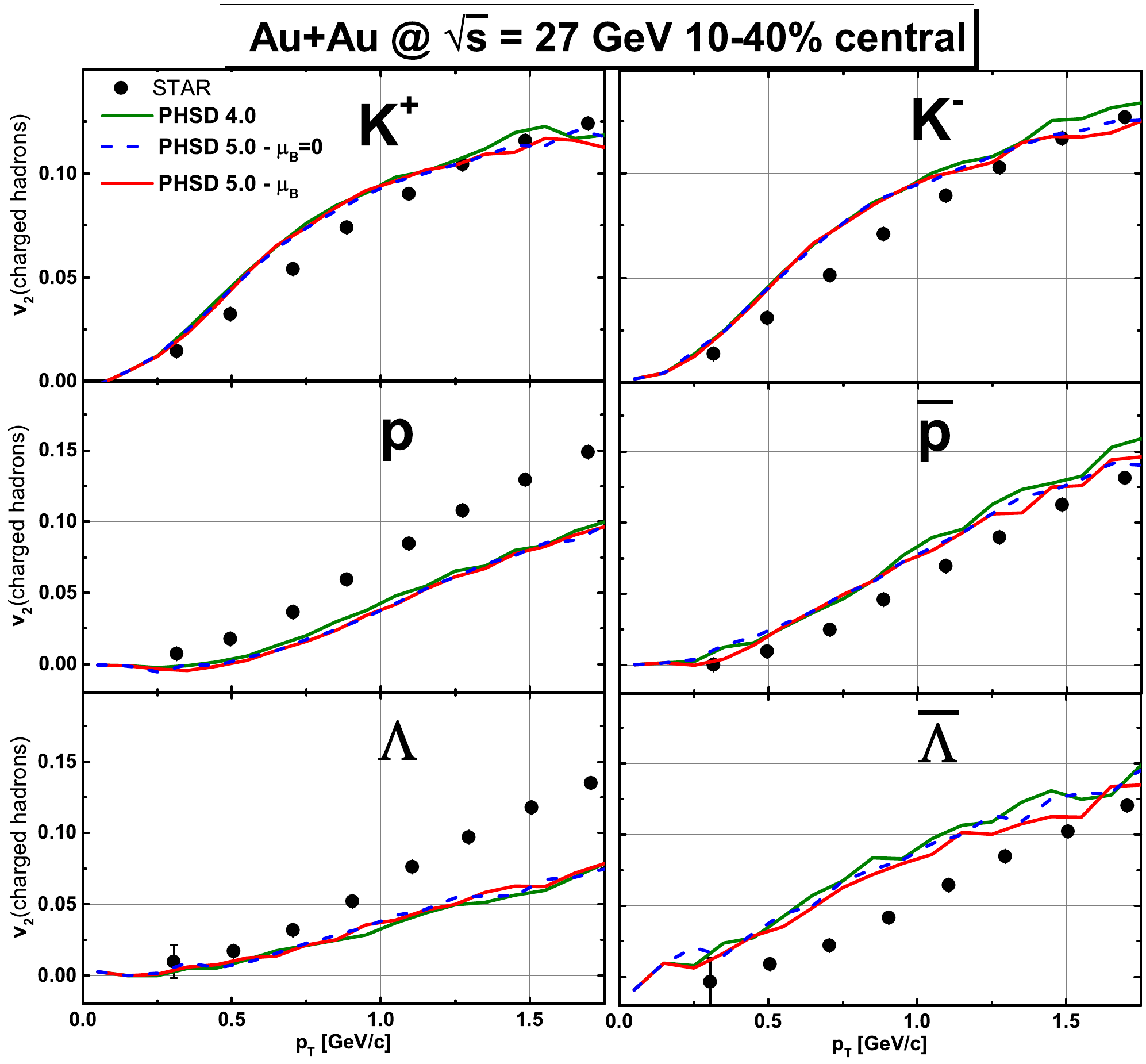} 
\caption{Elliptic flow of identified hadrons ($K^\pm, p, \bar p, \Lambda+\Sigma^0, \bar\Lambda + \bar\Sigma^0$) as a function of $p_T$ at $\sqrt{s_{NN}}$ = 27~GeV  for PHSD4.0 (green lines), PHSD5.0 with partonic cross sections and parton masses calculated for $\mu_B$ = 0 (blue dashed lines) and with cross sections and parton masses evaluated at the actual chemical potential $\mu_B$ in each individual space-time cell (red lines) 
in comparison to the experimental data of the STAR Collaboration~\cite{STARBESv2}.}
\label{v2y27GeV}
\end{figure}

In Figure~\ref{v2y27GeV}, we present the elliptic flow of identified hadrons ($K^\pm, p, \bar p, \Lambda+\Sigma^0, \bar\Lambda + \bar\Sigma^0$) as a function of  $p_T$ at 
$\sqrt{s_{NN}}$ = 27 GeV  for PHSD4.0 (green lines), PHSD5.0 with partonic cross sections and parton masses calculated for $\mu_B$ = 0 (blue dashed lines) and with cross sections and parton masses evaluated at the actual chemical potential $\mu_B$ in each individual space-time cell (red lines) in comparison to the experimental data of the STAR Collaboration~\cite{STARBESv2}. 
Similar to the directed flow shown in Figure~\ref{v1y27GeV}, the~elliptic flow from all 
three cases for PHSD shows a rather similar behavior,  the~differences are very small (within the statistics achieved here). Only antiprotons and antihyperons show a small decrease of  
$v_2$ at larger $p_T$ for PHSD5.0 compared to PHSD4.0, which can be attributed
to the explicit $\sqrt{s}$-dependence and different angular distribution of partonic
cross sections in the PHSD5.0. We note that the underestimation of $v_2$ for protons
and $\Lambda$'s we attribute to the details of the hadronic vector potential involved
in this calculations which seems to underestimate~repulsion.

\section{Conclusions}

In this work, we have studied the influence of the baryon chemical potential $\mu_B$ 
on the properties of the QGP in equilibrium as well as the QGP created in heavy-ion collisions 
also far from~equilibrium.

For the description of the QGP, we employed the extended effective
Dynamical QuasiParticle Model (DQPM) that is matched to reproduce the lQCD crossover
equation-of-state versus temperature $T$ and at finite baryon chemical potential $\mu_B$.
We compared the DQPM results for transport coefficients such as shear viscosity $\eta$ 
and bulk viscosity $\zeta$ with available lQCD data and the non-conformal 
holographic model at $\mu_B=0$  
and with results from a Bayesian analysis of experimental heavy-ion data.
We find that the ratios $\eta/s$ and $\zeta/s$ from the DQPM agree very well with 
the lQCD results from Ref.~\cite{Astrakhantsev:2018oue} and show a similar behavior
as the ratio obtained from a Bayesian fit~\cite{Bernhard:2016tnd}. As~found in~\cite{Moreau:2019vhw,Soloveva:2019xph}, the~transport coefficients 
show a mild dependence on $\mu_B$.

Following~\cite{Moreau:2019vhw}, we based our study of the non-equilibrium QGP---as created 
in heavy-ion collisions---on the extended Parton--Hadron--String Dynamics (PHSD) 
transport approach in which i) the masses and widths of quarks and gluons depend on $T$
and $\mu_B$ explicitly; ii) the partonic interaction cross sections are obtained 
by calculations of the leading order Feynman diagrams from the DQPM and explicitly depend on 
the invariant energy $\sqrt{s}$, temperature $T$ and baryon chemical potential $\mu_B$. 
This~extension is realized in the full version of PHSD5.0~\cite{Moreau:2019vhw}.

In order to investigate the traces of the $\mu_B$ dependence of the QGP in observables, 
the results of PHSD5.0 with $\mu_B$ dependences have been compared to the
results of PHSD5.0 for $\mu_B=0$ as well as with PHSD4.0 where 
the masses/width of quarks and gluons as well as their interaction cross sections
depend only on $T$ following Ref.~\cite{Ozvenchuk13}.
We have presented the PHSD results for different observables:
(i)~rapidity and $p_T$ distributions of identified hadrons for asymmetric Cu+Au 
collisions at energies of 30 AGeV (future NICA energy) as well as for the top RHIC energy
of  $\sqrt{s_{NN}}=200$ GeV;
(ii) directed flow $v_1$ of identified hadrons for $Au+Au$ at invariant 
energy $\sqrt{s_{NN}}=27$ GeV;
(iii) elliptic flow  $v_2$ of identified hadrons for $Au+Au$ at invariant 
energies $\sqrt{s_{NN}}=27$ and 200 GeV.
We find only small differences between PHSD4.0 and PHSD5.0 results on the hadronic observables
considered here at high as well as at intermediate energies. 
This is related to the fact that at high energies, where the matter is dominated by the QGP, 
one probes a very small baryon chemical potential in central collisions at midrapidity,
while, with decreasing energy, where $\mu_B$ becomes larger, the~fraction of the QGP drops rapidly, such that in total the final observables are
dominated by the hadrons which participated in hadronic rescattering and thus the~information about their QGP origin is washed out.
We have shown that the $\mu_B$ dependence of QGP interactions is more 
pronounced in observables for strange hadrons, kaons and especially 
anti-strange hyperons, as~well as for antiprotons.  
This gives an experimental hint for the searching of  $\mu_B$
traces of the QGP for experiments at the future NICA accelerator, even if it will 
be a very challenging experimental~task.

\vspace{6pt}

\acknowledgments{The authors acknowledge inspiring discussions with J. Aichelin, W. Cassing, 
V. Kolesnikov,  I.  Selyuzhenkov and A. Taranenko. We thank M. Attems 
for providing us the
results from Ref. \cite{Mykhaylova:2019} in data form.
Furthermore, we acknowledge support by the Deutsche Forschungsgemeinschaft (DFG, German Research Foundation) through the grant CRC-TR 211 'Strong-interaction matter under extreme conditions' - Project number 315477589 - TRR 211. O.S. acknowledges support from HGS-HIRe for FAIR; L.O. and E.B. thank the COST Action THOR, CA15213.
L.O. has been financially supported in part by the Alexander von Humboldt Foundation.
The~computational resources have been provided by the LOEWE-Center for Scientific~Computing.}

\conflictsofinterest{The authors declare no conflict of interest.}

\reftitle{References}


\end{document}